# Improving the Stability of Colloidal CsPbBr$_3$ Nanocrystals with an Alkylphosphonium Bromide as Surface Ligand Pair

Meenakshi Pegu[1,#], Hossein Roshan[2,#], Clara Otero-Martínez[1], Luca Goldoni[3], Juliette Zito[1], Nikolaos Livakas[1,4], Pascal Rusch[1], Francesco De Boni[3], Francesco Di Stasio[2], Ivan Infante,[5,6] Luca De Trizio[7]*, Liberato Manna[1]*

[1] Nanochemistry, [2] Photonic Nanomaterials, [3] Materials Characterization, [7] Chemistry Facility, Istituto Italiano di Tecnologia, Via Morego 30, 16163 Genova, Italy

[4] Dipartimento di Chimica e Chimica Industriale, Università di Genova,16146 Genova, Italy

[5] BCMaterials, Basque Center for Materials, Applications, and Nanostructures, UPV/EHU Science Park, Leioa 48940, Spain

[6] Ikerbasque Basque Foundation for Science, Bilbao 48009, Spain

AUTHOR INFORMATION
# These authors contributed equally to this work (M.P and H.R)
Corresponding Authors:
luca.detrizio@iit.it;
liberato.manna@iit.it

**Abstract:**

In this study, we synthesized a phosphonium-based ligand, trimethyl(tetradecyl)phosphonium bromide (TTP-Br), and employed it in the post-synthesis surface treatment of Cs-oleate-capped CsPbBr$_3$ NCs. The photoluminescence quantum yield (PLQY) of the NCs increased from ~60% to more than 90%, as a consequence of replacing Cs-oleate with TTP-Br ligand pairs. Density functional theory calculations revealed that TTP$^+$ ions bind to the NC surface by occupying Cs$^+$ surface sites and orienting one of their P−CH$_3$ bonds perpendicular to the surface, akin to quaternary ammonium passivation. Importantly, TTP-Br-capped NCs exhibited higher stability in air compared to didodecyldimethylammonium bromide-capped CsPbBr$_3$ NCs (which is considered a benchmark system), retaining ~90% of their PLQY after six weeks of air exposure. Light-emitting diodes fabricated with TTP-Br-capped NCs achieved a maximum external quantum efficiency of 17.2%, demonstrating the potential of phosphonium-based molecules as surface ligands for CsPbBr$_3$ NCs in optoelectronic applications.

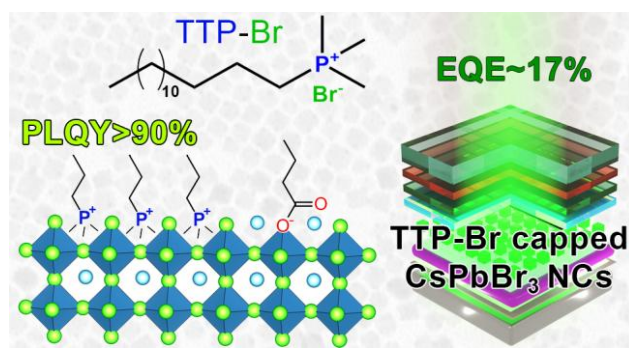

Colloidal nanocrystals (NCs) of lead halide perovskites, with chemical formula CsPbX$_3$ (X=Cl, Br, I), have emerged as promising active elements for optoelectronic devices, including light-emitting diodes (LEDs), displays, scintillators, solar concentrators, photodetectors, and solar cells.[1-3] Such interest stems from the remarkable optical properties of these NCs, which include high photoluminescence (PL) quantum yield (QY), narrow PL line widths, and tuneable optical band gaps that extend from the visible spectral region up to the near-infrared.[4-10] These NCs, however, suffer from two main drawbacks due to their strong ionic character: (i) their inherent high solubility in polar solvents makes them prone to partial dissolution or etching when exposed to moisture, air, or during standard purification procedures, and (ii) the surface ligands are typically weakly bound and tend to desorb from the surface. The latter point is critical, since even a partial desorption of surface ligands not only negatively affects the colloidal stability of CsPbX$_3$ NCs but also contributes to the degradation of their optical properties.[11-13]

In this context, typical surface ligands used in the synthesis of CsPbX$_3$ NCs are alkylamines and/or carboxylic acids, which bind to the NCs' surface as ion pairs, specifically as alkylammonium-halides and Cs-carboxylates,[14, 15] with alkylammonium cations occupying surface Cs$^+$ sites and carboxylate anions replacing surface halide anions. These charged ligands can easily detach from the surface of CsPbX$_3$ NCs upon protonation or deprotonation, for example by simply exposing the colloidal solution to air, eventually causing degradation in PLQY.[13, 16] To address these issues and improve both the colloidal stability and PLQY of CsPbX$_3$ NCs, various alternative surface ligands have been explored so far. The choice usually falls on molecules such as alkyl sulfonates or phosphonates[17, 18] that have a higher binding strength to the surface of the NCs compared to alkylammonium-halide and Cs-carboxylate, or molecules that are unaffected by protonation or deprotonation, such as alkyl quaternary ammonium halides or zwitterionic molecules.[16, 19-22] We cite here only some representative examples: Cai et al. successfully employed several alkyl sulfonium bromides for the colloidal synthesis of CsPbBr$_3$ NCs,[23] while Imran et al. replaced Cs-oleate ion couples on the surface of CsPbBr$_3$ NCs with didodecyldimethylammonium bromide (DDA-Br).[16] Kreig et al. reported an effective approach to synthesize CsPbBr$_3$ NCs using zwitterionic long-chain molecules such as sulfobetaine, phosphocholine, or γ-amino acid.[21] Two common features of CsPbBr$_3$ NCs from all these works are a near-unity PLQY and improved stability under ambient conditions and even upon washing with solvents to remove excess ligands.

Another unexplored class of ligands with the potential to deliver efficient and stable CsPbBr$_3$ NCs and deserving further investigation are the alkylphosphonium salts. Large aromatic phosphonium halide ion pairs have been used to treat CsPbX$_3$ NC films, enabling effective charge injection and mobility and thereby improving the efficiency of the corresponding perovskite-based LED (**Table S1**).[24-26] Despite these advancements, no in-depth investigation has yet been conducted to determine whether these species can efficiently bind to the surface of CsPbBr$_3$ NCs. Given the structural similarity between quaternary phosphonium halides and quaternary ammonium halides, we hypothesize that the quaternary phosphonium halides as well should also be capable of anchoring to the surface of CsPbBr$_3$ NCs without altering the NCs' morphology and potentially resulting in stable and strongly emissive NCs. To test this hypothesis, we synthesized trimethyl(tetradecyl)phosphonium bromide (TTP-Br), a quaternary alkyl phosphonium halide salt in which the four organic substituents bound to the phosphorous atom are three methyl groups and a long tetradecyl group, and evaluated its effectiveness as a surface ligand for CsPbBr$_3$ NCs.[27] This compound, which has the advantage of a relatively accessible phosphonium group while at the same time being soluble in most common organic solvents, was employed in a post-synthesis ligand exchange procedure involving Cs-oleate-capped CsPbBr$_3$ NCs in toluene (**Scheme 1**).

**Scheme 1. Sketch of the post-synthesis ligand exchange process involving Cs-Oleate-capped CsPbBr$_3$ NCs with TTP-Br.**

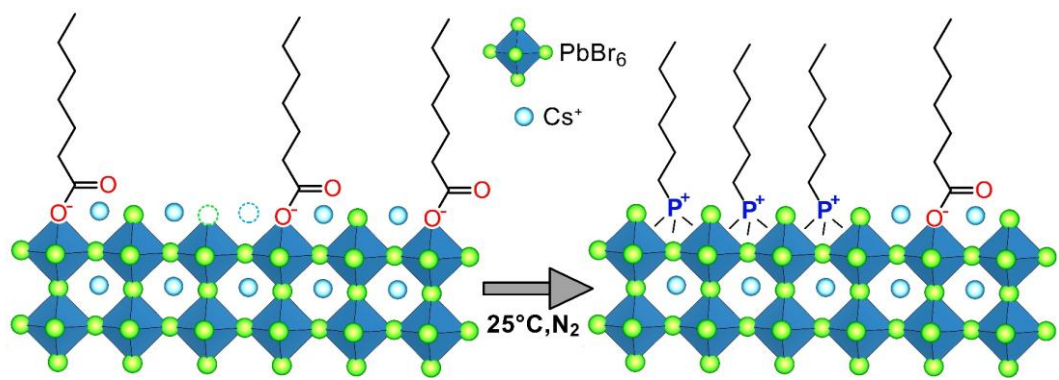

Our results demonstrated a successful exchange of Cs-oleate for TTP-Br, with the PLQY of $CsPbBr_3$ NCs increasing from 62% (as-synthesized NCs, coated with Cs-oleate) to over 90% (TTP-Br coated NCs). Nuclear magnetic resonance (NMR) analyses revealed that the ligand shell of the final NCs comprised 92% TTP-Br (with a density of 1.28 ligands/nm$^2$) and a residual 8% of Cs-oleate. To rationalize these findings, we performed density functional theory (DFT) calculations, which revealed that TTP-Br has a high binding energy to the surface of $CsPbBr_3$ NCs (42.73 kcal/mol), a value comparable to that of DDA-Br.[11]

To evaluate whether this quaternary phosphonium salt could provide better surface passivation for $CsPbBr_3$ NCs compared to quaternary alkylammonium-based salts, we also performed the same ligand exchange procedure using the corresponding ammonium bromide pair, that is, trimethyl(tetradecyl)ammonium bromide (TTN-Br). Surprisingly, ligand exchange with TTN-Br led to the precipitation of $CsPbBr_3$ NCs, which could not be redispersed in any organic solvent. Therefore, we decided to compare the stability of TTP-Br-capped $CsPbBr_3$ NCs to that of DDA-Br-capped NCs, which are widely regarded as a benchmark system due to their near unity PLQY and high colloidal stability under various ambient stimuli such as heat, and solvent washing, and upon long-term storage.[16, 28] Additionally, the ligand exchange procedure that we employed to prepare TTP-Br-capped NCs is analogous to the one reported to prepare DDA-Br-capped NCs, with the starting point being Cs-oleate-capped $CsPbBr_3$ NCs in both cases.[16] This allows for a clear comparison between the two ligands, as the two end products feature the same inorganic core but different ligand shells.

The comparison was done by exposing both NC solutions to air for a time span of six weeks, during which the TTP-Br-capped NCs exhibited higher stability compared to DDA-Br-capped NCs, retaining ~90% of their initial PLQY at the end of this time span, while the DDA-Br-capped NCs retained only ~76% of their initial PLQY. The higher air stability of the TTP-Br-capped NCs, along with the fact that TTP-Br ligands are less bulky and, therefore, likely more electrically conductive than DDA-Br (the former has two dodecyl alkyl chains, while the latter has only one tetradecyl alkyl chain), motivated us to fabricate green-emitting LED devices with both TTP-Br-capped and DDA-Br capped NCs and make a comprehensive comparison. The LEDs with TTP-Br capped NCs achieved a maximum external quantum efficiency (EQE) of 17.2 % at high luminance of 2600 cd m$^{-2}$, significantly surpassing the EQE of DDA-Br-based LED with 7.4% EQE. Moreover, stability tests under operation conditions showed higher operational stability of TTP-Br LEDs in comparison with the DDA-Br counterparts.

We synthesized TTP-Br using a one-pot synthesis approach, which involves the nucleophilic substitution reaction of trimethylphosphine with tetradecyl halide (Equation 1, **Scheme S1**) with a reaction yield of ~75% (see the Experimental Section, **Figure S1-S3** and **Scheme S1** of the Supporting Information, SI, for details).

$$(CH_3)_3P + C_{14}H_{29}\text{-Br} \rightarrow C_{14}H_{29}(CH_3)_3P^+\text{-Br}^- \qquad (Eq.\ 1)$$

The purified TTP-Br salt is soluble at room temperature in various organic solvents, including chloroform and partially in toluene, therefore it could be readily employed in a ligand exchange procedure involving Cs-oleate-capped CsPbBr$_3$ NCs. Cs-oleate-capped CsPbBr$_3$ NCs were synthesized following the well-established hot injection protocol reported by Imran et al. with minor modifications (see the Experimental Section).[16, 29] The ligand exchange process was performed as follows: after quenching the synthesis of Cs-oleate-capped NCs, 2 mL of a 25 mM solution of TTP-Br in chloroform/toluene (10% v/v) was added to a portion of the crude reaction solution (3 mL) containing 0.09 mmol of CsPbBr$_3$ NCs. The reaction was allowed to proceed for 20 minutes under a nitrogen atmosphere, after which the product was washed with ethyl acetate and re-dispersed in toluene (see the Experimental Section for more details).

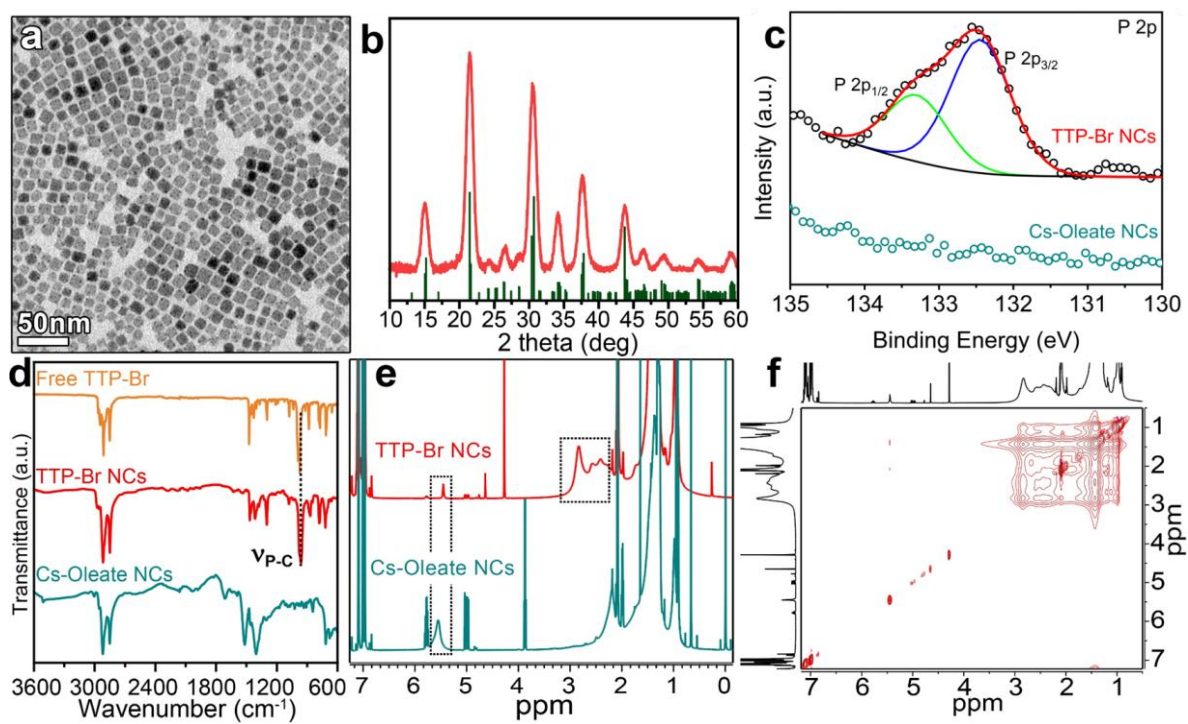

**Figure 1**. (a) TEM micrograph of TTP-Br-capped CsPbBr$_3$ NCs. (b) X-ray diffraction pattern of TTP-Br-capped CsPbBr$_3$ NCs and reference CsPbBr$_3$ orthorhombic bulk reflections (ICSD number 243735). (c) XPS P *2p* spectra of TTP-Br- and Cs-Oleate-capped CsPbBr$_3$ NCs. (d) FTIR spectra of TTP-Br ligand, TTP-Br- and Cs-Oleate-capped CsPbBr$_3$ NCs. (e) $^1$H NMR spectra at 298K of Cs-Oleate- and TTP-Br-capped CsPbBr$_3$ NCs, the full $^1$H spectrum of the latter with peaks assignment is reported (see the Supporting Information for more details). (f) 2D NOESY spectrum of TTP-Br-capped CsPbBr$_3$ NCs at 313K (see the Supporting Information for the detailed spectrum).

Upon ligand exchange with TTP-Br, the NCs retained their cubic shape, with a slight size reduction from 9.7 ± 1.2 nm (**Figure S4a**) to 8.9 ± 1.6 nm (**Figure 1a**), indicating a minor etching of the NCs, compatible with what reported for the DDA-Br case.[16] X-ray diffraction (XRD) patterns of the CsPbBr$_3$ NCs before and after the exchange featured the characteristic peaks of the CsPbBr$_3$ orthorhombic phase (ICSD number 243735) at 15.05º, 21.54º, and 30.56º, corresponding to the (100), (110), and (200) lattice planes, respectively (**Figure 1b** and **Figure S4b**). To further investigate the effects of the ligand exchange procedure, we performed X-ray photoelectron spectroscopy (XPS) measurements. Upon exchange with TTP-Br, the relative atomic percentage of O decreased from 9.3% to 2.8%, while the Br/Pb atomic ratio increased from 2.28 to 2.75 (**Table S2, Figures S5-S6**). Moreover, the final sample exhibited additional XPS peaks at 132.6 eV and 133.4 eV, ascribed to P *2p*, corroborating the effective anchoring of TTP-Br onto the NCs' surface (**Figure 1c**). The presence of homogeneously distributed P on the TTP-Br-exchanged CsPbBr$_3$ NCs was further confirmed by transmission electron microscopy (TEM) and scanning electron microscopy (SEM) energy-dispersive X-ray spectroscopy (EDX) mapping (**Figure S7-S8, Tables S3-S4**).

To assess the degree of the Cs-oleate→TTP-Br exchange and to reveal the ligand shell composition of the exchanged NCs, we performed both Fourier-transform Infrared (FTIR) and NMR analyses. The FTIR spectra of the NCs before and after the TTP-Br exchange revealed a significant replacement of oleate species with TTP-Br: i) a reduction in the peaks at 1710 cm$^{-1}$ and 1535 cm$^{-1}$ corresponding to the asymmetric and symmetric stretching vibrations of the C=O and COO$^-$ groups, respectively (**Figure S9**); ii) the emergence of a peak at approximately 990 cm$^{-1}$, attributable to the P-C stretching vibrations (**Figure 1d**).[27, 30] Similarly, the $^1$H NMR spectra of CsPbBr$_3$ NCs before and after ligand exchange with TTP-Br showed a reduction of the signal at 5.48 ppm, ascribed to the protons of the double bond in the oleate species, along with the appearance of broad peaks between 2.01 and 2.84 ppm (**Figure 1e**). Such peaks were attributed to the Me$_3$-P groups of TTP-Br bound to the surface of the NCs (see **Figures S1-S3** and **Figures S10-S12** for complete signals assignment of the TTP-Br ligand and TTP-Br-capped CsPbBr$_3$ NCs). At 298 K, the free TTP-Br ligand was observed to form micelles or aggregates in toluene-D, as indicated by the $^1$H-$^1$H NOESY spectrum (**Figure S13**). To ensure complete solubility of TTP-Br and accurate assignment of the peaks on the NCs' surface, NMR measurements of the TTP-Br ligand as well as TTP-Br-capped NCs were conducted at 313K (**Figure S14-S16**). The dynamic interaction of the TTP-Br ligand with the NCs' surface was demonstrated by $^1$H-$^1$H NOESY at 313K,[19] which evidenced negative (red) NOE cross peaks for the TTP-Br capped NCs (**Figure 1f**), typical of species with a long correlation time ($\tau_c$), i.e. with a slow tumbling regime in solution. Positive (blue) NOE cross-peaks were observed for the free TTP-Br at 313K, characteristic of a molecule with the fast-tumbling regime in solution (**Figure S15**). To quantify the extent of Cs-oleate→TTP-Br replacement, we performed a quantitative NMR (*q*-NMR) analysis of the final NC sample dissolved in deuterated dimethyl sulfoxide (**Figure S17**), coupled with elemental analysis of the same solution performed via inductively coupled plasma optical emission spectroscopy (ICP-OES).[31, 32] Our results indicate that 92% of the ligand shell was composed of TTP-Br**,** while the remaining 8% consisted of Cs-oleate, with a calculated surface density of 1.28 TTP-Br molecules per nm$^2$ and 0.10 oleate species per nm$^2$ (see **Table S5** in the Supporting Information for details).

In order to assess the potential advantages of using TTP-Br compared to quaternary alkyl ammonium salts in the passivation of $CsPbBr_3$ NCs, we carried out the same ligand exchange procedure using trimethyl(tetradecyl)ammonium bromide (TTN-Br), the ammonium analogue of TTP-Br. Unfortunately, the NCs precipitated during the exchange and the collected product could not be re-dispersed in toluene nor in chloroform. Therefore, we decided to compare the TTP-Br-capped NCs with DDA-Br-capped ones, the latter representing one of the best $CsPbBr_3$ NC systems regarding PLQY and colloidal/air stability.[11] DDA-Br-capped NCs were prepared using the same ligand exchange protocol employed for TTP-Br-capped NCs, starting from Cs-oleate-capped NCs (see the Experimental Section for details). The resulting DDA-Br-capped NCs had a size of 8.8 ± 1.9 nm, with XRD patterns corresponding to the orthorhombic phase of the $CsPbBr_3$ NCs (**Figure S18**). Similar to our previous reports, XPS and FTIR analyses confirmed the surface binding of the DDA-Br ligand on the NCs (**Table S2, Figures S19-S20**).[16] The NMR measurements confirmed the effective binding of DDA-Br ligand on the NCs' surface (**Figures S21-S25**), with DDA-Br and Cs-oleate coverage of 1.32 and 0.27 ligands/$nm^2$, respectively (**Table S5**, **Figure S26**).[16, 29]

In terms of optical properties, both TTP-Br- and DDA-Br-capped NCs exhibited a slightly blue-shifted PL and excitonic absorption peaks with respect to the initial Cs-oleate-capped NCs (**Table 1**, **Figure 2a**), consistent with the slight reduction in size observed after the exchange in both cases. The PLQY increased from 62 ± 6% to 91 ± 9% in the TTP-Br case and to 98 ± 9% in the DDA-Br case (**Table 1**). PL decay measurements revealed that the PL lifetime ($\tau_{avg}$) increased in solution for both the DDA-Br and TTP-Br samples, from 8.86 ns (for Cs-oleate-capped NCs) to 10.74 ns and 11.29 ns, respectively (**Figure 2b**, **Figure S27, and Table S6**). As these increases in PL lifetime go hand in hand with increases in PLQY, we conclude that fast, non-radiative recombination is significantly reduced in these ligands exchanged NCs, most likely due to more effective surface passivation. PL measurements performed in thin films at temperatures ranging from 7 K to 267K indicated that both DDA-Br-capped and TTP-Br-capped NCs featured the typical temperature-dependent behavior of lead halide perovskite NCs, that is, i) a blue shift of the PL emission and ii) an increase in PL lifetime when raising the temperature from 7K to 267K (see **Figure S28** and **Tables S7-S8** of the Supporting Information for details).[33, 34] The temperature-dependent blue-shift is attributed to the thermal expansion of the perovskite lattice,[33] which reduces the overlap between the Pb 6s and Br 4p orbitals and results in a lower energy of the antibonding crystal orbital at the valence band maximum, leading to an increase in the bandgap energy.[34, 35] The shorter average lifetime at lower temperatures is consistent with the exciton ground state being a bright triplet for NCs of sizes around 9-10 nm (as those studied in this work) or beyond, in agreement with previous experimental and theoretical works.[10, 36, 37]

To evaluate the stability of all the samples, we exposed NCs dispersions in toluene to air for a period of six weeks. Among the three samples, TTP-Br-capped NCs exhibited the best stability featuring a PLQY of ~85% at the end of the test, while DDA-Br- and Cs-oleate-capped NCs exhibited a PLQY of ~76 and ~3%, respectively (**Figure 2d**). Both the TTP-Br- and DDA-Br capped NCs were observed to slightly increase in size from 9.8 ± 1.8 nm to 10.3 ± 1.6 nm after six weeks of exposure to air (**Figure S29**), likely due to a slow Oswald ripening process. XRD patterns of DDA-Br and TTP-Br-capped NCs exposed to air after six weeks were compatible with the presence of the orthorhombic phase of $CsPbBr_3$ crystal structure, with no evidence of secondary phases, indicating good structural stability in both samples (**Figure S30**). Quaternary phosphonium/ammonium ions have a stable positive charge, unlike, for example, primary ammonium or carboxylate ions, which can be instead deprotonated/protonated upon exposure to air (operated by humidity/water present in the air). This prevents these ligands from losing their charge, allowing them to continue binding to the surface of the NCs even upon exposure to air. We believe that this property enables these ligands to sufficiently protect the NCs from water/humidity, which is known to trigger the phase change from $CsPbBr_3$ to $Cs_4PbBr_6$.[12, 16, 38] Indeed, even after extending their exposure to air for up to a period of two months and a half, both the TTP-Br-capped and DDA-Br-capped $CsPbBr_3$ NCs did not transform, not even partially, into $Cs_4PbBr_6$, as proven by XRD analyses (**Figure S31**).

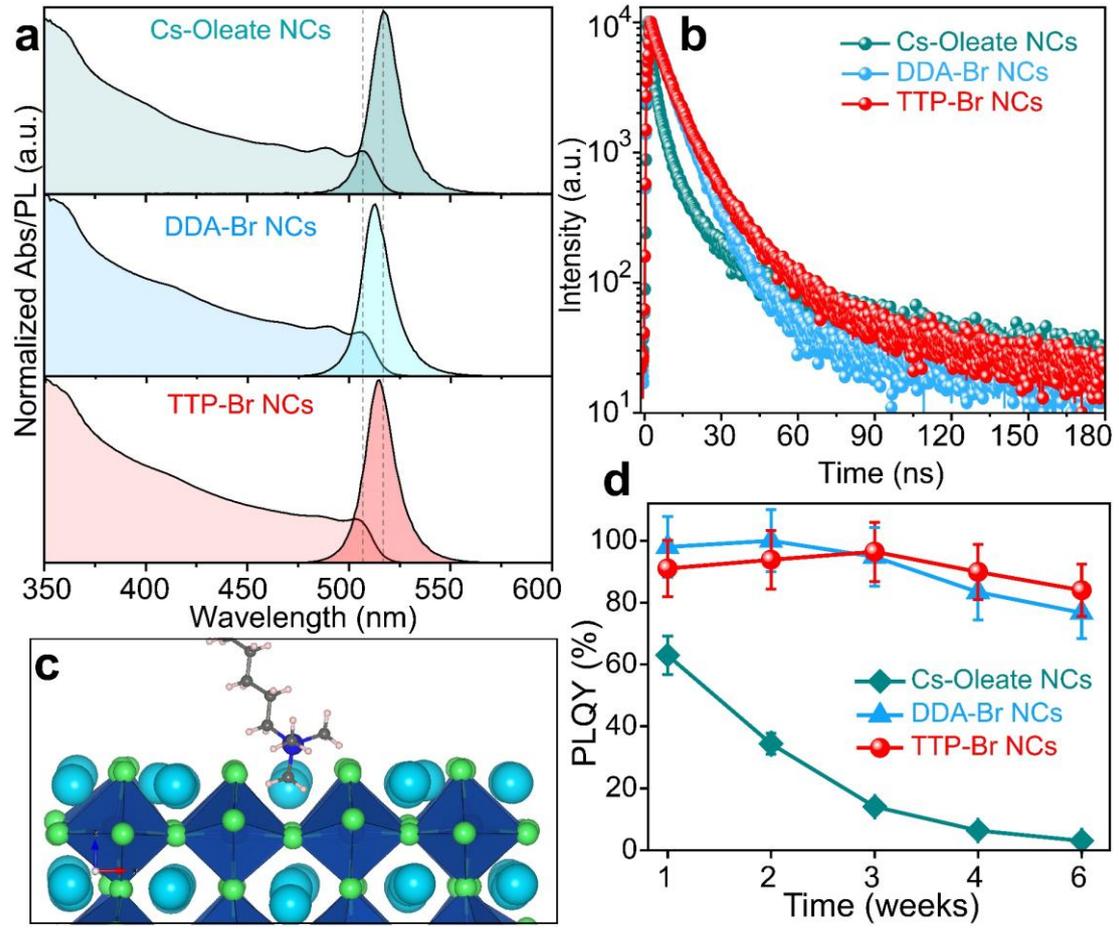

**Figure 2.** (a) UV-visible absorption and PL spectra of Cs-Oleate-, DDA-Br-, and TTP-Br-capped CsPbBr$_3$ NCs. (b) PL decay profile of Cs-Oleate-, DDA-Br-, and TTP-Br- capped CsPbBr$_3$ NCs. (c) Binding configuration of TTP-Br- ligands sitting in the A-site of the CsPbBr$_3$ NCs' surface. (d) PLQY stability of Cs-Oleate-, DDA-Br-, and TTP-Br-capped CsPbBr$_3$ NCs over time at ambient storage conditions.

**Table 1. Optical properties of the different NC samples in solution at room temperature.**

| NCs Sample | Abs$_{max}$ (nm) | PL$_{max}$ (nm) | FWHM (nm) | PLQY (%) Week 1 | PLQY (%) Week 2 | $\tau_{avg}$ (ns) |
|---|---|---|---|---|---|---|
| **Cs-Oleate** | 508 | 517 | 18.5 | 62±6 | 36±4 | 8.86 |
| **DDA-Br** | 505 | 512 | 17.2 | ~98 | ~100 | 10.74 |
| **TTP-Br** | 506 | 514 | 18.0 | 91±9 | 96±9 | 11.29 |

To better understand how phosphonium ligands bind to the surface of CsPbBr$_3$ NCs and whether this differs significantly from quaternary ammonium ligands, we performed DFT calculations (details in the Experimental Section). We first examined the binding characteristics of a single ion pair on the NC surface. To do this, we constructed a cubic, charge-balanced CsPbBr$_3$ NC model with a ~2.4 nm edge-size and replaced one CsBr unit from the outer AX shell with either a TTP-Br or DDA-Br ion pair. Upon structural relaxation, we observed that the phosphonium group oriented one of its P−CH$_3$ bonds perpendicular to the NC surface, similar to the orientation observed for the quaternary ammonium group (**Figures 2c** and **S32**).[16] The calculated binding energies for TTP-Br (42.7 kcal/mol) and DDA-Br (45.2 kcal/mol) were also comparable and in line with reported values for primary alkylammonium-Br (45.3 kcal/mol), secondary alkylammonium-Br (48.2 kcal/mol), and zwitterionic ligands such as sulfobetaine (41.0 kcal/mol), at similar theoretical levels of calculations.[21,39] This consistency has been ascribed to the dominance of electrostatic interactions in determining the binding strength with the perovskite NC surface, regardless of the type of anchoring group used. To investigate the enhanced PL and stability of NCs passivated by TTP-Br ligands, we replaced CsBr units on all six facets with alkyl phosphonium bromide pairs, achieving a surface concentration of 1.27 ligands/nm²—consistent with the experimental 1.28 ligands/nm². To reduce system size and computational cost, we simplified TTP-Br by replacing the tetradecyl group with ethyl. As shown in **Figure S33**, the electronic structure remains free of midgap states, with a fully delocalized valence band maximum, similar to quaternary ammonium passivation.[16] In contrast, purely Cs-oleate-capped NCs, as demonstrated by earlier DFT calculations[19], exhibit trap states at both high and low ligand concentrations, localized on the oxygen atoms in the carboxylate anchoring groups. This enables nonradiative recombination and contributes to the suboptimal PL QY observed in experiments. Ligand exchange with phosphonium or quaternary ammonium pairs effectively eliminates oxygen-related defects.

The improved air stability of TTP-Br-capped NCs, along with the fact that TTP-Br is less bulky and likely less electrically resistive than DDA-Br, could lead to improved charge injection. This motivated us to test these NCs in optoelectronic devices, specifically LEDs. The LEDs were prepared using a conventional structure in which the emissive layer is sandwiched between a hole transport layer (HTL) and an electron transport layer (ETL), with the HTL being deposited first. As HTL layers, we tested poly(triaryl amine) (PTAA) and polyvinyl carbazole (PVK), while as the ETL, we employed 2′,2′-(1,3,5-benzinetriyl)-tris(1-phenyl-1-H-benzimidazole) (TPBi). We compared our devices with a double HTL (PTAA/PVK) architecture, namely

ITO/PEDOT:PSS/PTAA/PVK/CsPbBr$_3$ NCs/TPBi/LiF/Aluminum (**Figure 3a**), to those with a single HTL using either PTAA or PVK. Here, PEDOT is poly(3,4-ethylenedioxythiophene), PSS is polystyrene sulfonate, and LiF is lithium fluoride. A statistical EQE study demonstrated the significant superiority of double HTL LEDs over those with single HTL (**Figure S34**), showing that the double HTL architecture can achieve optimal band alignment and efficient hole injection into the emissive layer, thereby improving the charge balance in the active layer.[40] **Figure 3b** illustrates the band alignment of the emitting layer, as determined by UPS analysis and optical absorption spectra (**Table S9, Figure S35-S36**). It also depicts the energy levels of all other layers of the double HTL architecture, as well as those of the electrodes. The energy level values of ITO, PEDOT:PSS, PTAA, PVK, TPBi, and LiF/Aluminium layers (**Figure 3b**) were taken from our previous work.[41] The electroluminescence (EL) spectrum of the LED showed an emission peak at 516 nm with a full width at half maximum (FWHM) of 18 nm (~84 meV), consistent across all bias voltages (**Figure 3c**). This matched the PL spectrum of the solid TTP-Br-capped NC film (PL peak at 514 with FWHM of ~17.7 nm; thin film images of TTP-Br-capped and DDA-Br-capped NCs are presented in **Figure S37-S38**), with only a minor red shift observed in the peak position of EL spectra (**Figure 3c**). No other EL peaks were observed, indicating the absence of deep traps or interlayer emission. The 1931 CIE diagram of the EL spectrum, obtained at a 6 V operating voltage (**Figure S39**), showed that the LED had CIE coordinates (x = 0.084, y = 0.775), indicating a high degree of saturation in the green region, which translated to a vibrant and intense green emission (see also inset of **Figure 3c**). The gamut coverage of the green LED fell well within the Rec.709 color space, making it interesting for integration into displays. The LEDs statistically exhibited a turn-on voltage between 2.6 to 3V, which is slightly higher than the bandgap energy of the emissive layer (CsPbBr$_3$ NCs). The champion device demonstrated a luminance exceeding 10000 cd/m² (**Figure 3d**) at a current density of 43 mA/cm$^2$. The maximum EQE of the champion TTP-Br-capped NCs LED was 17.2%, and it was achieved at a luminance of 2600 cd/m², corresponding to an operation bias voltage of 4.3 V and a current density of 5.8 mA/cm$^2$ (**Figure 3e**). This relatively high EQE can be attributed to the high luminance of LEDs based on TTP-Br-capped NCs, which allowed efficient light emission while maintaining a relatively low current density.

The performance of the TTP-Br-capped NC LEDs was compared to that of the DDA-Br-capped NC LEDs that were prepared by employing DDA-Br-capped CsPbBr$_3$ NCs and using the same LED architecture incorporating a double HTL. The current density and luminance versus driving voltage curves for the double HTL-based DDA-Br-capped NC LED are presented in **Figure**

**S40**. The maximum luminance achieved was 1313 cd/m², while the EQE reached a maximum value of 7.4% (**Figure 3e)**, consistent with what was reported in a previous work from our group on similar LEDs.[41] It is worth highlighting that although our reported EQE is lower compared to the state-of-the-art value of 28.9%,[42] it represents a significant improvement over the highest EQE values reported for LED based on NCs capped by long-chain ligands (~15%, 9.8%, and 13.4%), specifically for DDA-Br-capped NCs.[28, 41, 43, 44] To investigate the origins of performance differences between TTP-Br-capped and DDA-Br-capped NC LEDs, we performed space-charge limited current (SCLC) measurements. The hole carrier mobility ($\mu$) was determined for both DDA-Br-capped and TTP-Br-capped NCs-based devices by fitting their dark J-V curves to the SCLC model (**Figure S41 a-b, Table S10**).[41, 42] The TTP-Br-capped NC-based device exhibited a higher hole mobility (1.43 x $10^{-6}$ cm²V⁻¹s⁻¹) compared to the DDA-Br-capped NCs (6.87 x $10^{-7}$ cm²V⁻¹s⁻¹). Moreover, the trap-filled limit voltage ($V_{TFL}$) of the device based on TTP-Br-capped NCs (0.27 V) was found to be lower than that of the device based on DDA-Br-capped NCs (0.36 V), suggesting that TTP-Br-capped NCs feature a lower density of surface traps and better surface passivation. Since both TTP-Br-capped and DDA-Br-capped samples exhibit comparable PLQY values, the superior charge transport in TTP-Br NCs likely contributes to the enhanced EQE observed in TTP-Br-based LEDs. The performance of the LEDs and the stability of the DDA-Br-capped and TTP-Br-capped NCs in solution motivated us to evaluate the luminance stability of the LEDs over time. To mitigate environmental effects, DDA-Br-capped and TTP-Br-capped NC LEDs were encapsulated with epoxy glue and tested under a constant applied voltage, corresponding to an initial luminance of 500 cd/m² (**Figure 3f).** After an initial rise, the luminance began to degrade, and the performance was assessed using the half-lifetime ($T_{50}$) criterion.[45] The initial increase in light intensity during stability tests may result from ion migration, defect passivation, and charge redistribution, leading to a transient enhancement in luminescence efficiency.[46] The TTP-Br-capped NCs LED exhibited a $T_{50}$ of 49 minutes, while the DDA-Br-capped NCs LED reached $T_{50}$ after 30 minutes. Furthermore, we examined the LEDs stability under high-luminance conditions by applying a bias voltage of 7 V and monitoring the luminance decay over time (inset of **Figure 3f)**. No significant differences between DDA-Br-capped and TTP-Br-capped NC LEDs were observed under high luminance, with both showing a $T_{50}$ of only a few minutes.

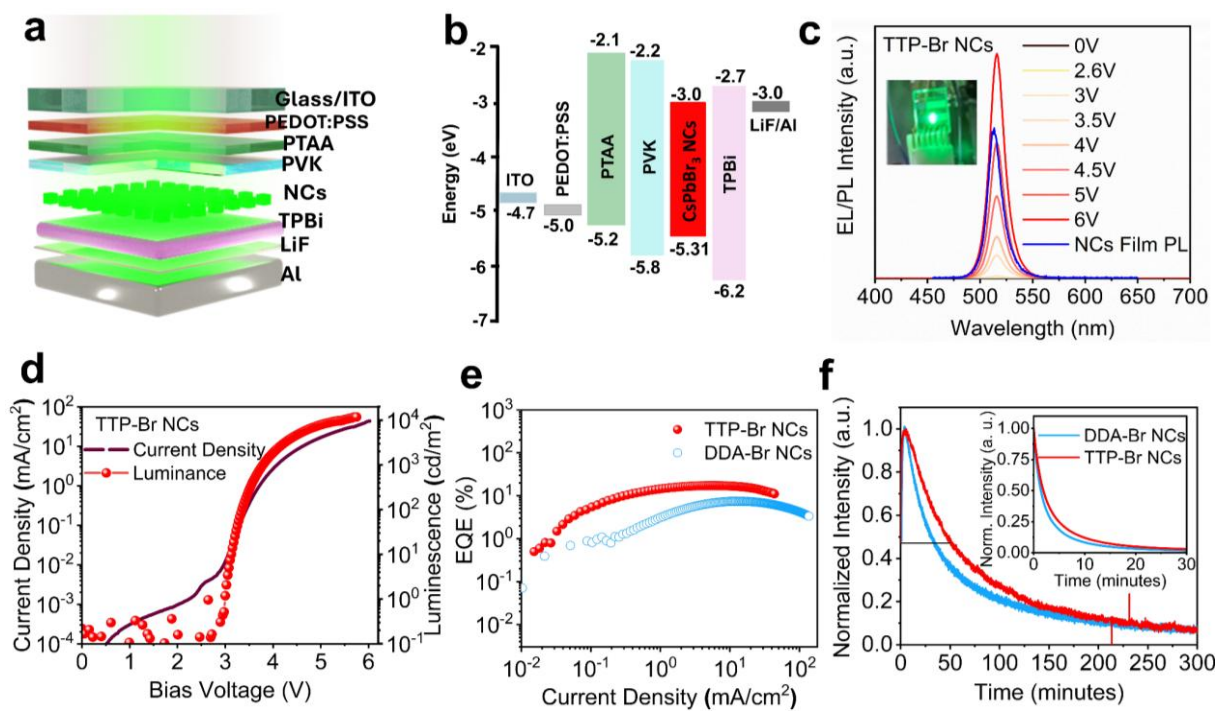

**Figure 3**. (a) Double HTL LED device configuration based on TTP-Br-capped CsPbBr$_3$ NCs. (b) Band-energy alignment of TTP-Br-capped CsPbBr$_3$ NCs with charge injection layers. (c) PL spectrum of TTP-Br-capped CsPbBr$_3$ NC film and EL spectra of the double HTL LED at different applied voltages. (d) Current density and luminance versus driving voltage curves of the double HTL TTP-Br-capped CsPbBr$_3$ NCs based champion LED device. (e) EQE versus current density of the double HTL LEDs champion devices based on TTP-Br-capped and DDA-Br-capped CsPbBr$_3$ NCs. (f) Stability of the EL intensity under operation conditions with the initial luminance of 500 cd/m$^2$ while the inset shows the intensity of light plotted over time with an applied bias of 7 V.

In summary, we prepared trimethyl(tetradecyl)phosphonium bromide (TTP-Br), a quaternary alkyl phosphonium halide salt, and tested it in a post-synthesis ligand exchange procedure involving colloidal Cs-oleate-capped CsPbBr$_3$ NCs. As a result of the Cs-oleate → TTP-Br exchange, the PLQY of CsPbBr$_3$ NCs increased from 62% to 91%. NMR analyses confirmed the successful ligand replacement, indicating a final ligand shell composed of 92% TTP-Br (with a density of 1.28 ligands/nm$^2$) and 8% Cs-oleate. The air stability of TTP-Br-capped NCs was compared to that of DDA-Br-capped NCs, which are considered a standard due to their high PLQY and air stability. After six weeks of air exposure, TTP-Br-capped NCs retained 90% of their PLQY, while the DDA-Br-capped NCs retained only 76%. DFT calculations revealed that TTP$^+$ ions bind to the NCs' surface by occupying A-sites and orienting one of their P−CH$_3$ bonds perpendicular to the NC surface. The calculated binding energy is 42.7 kcal/mol, comparable to that of DDA-Br or alkylammonium-Br species in general. The higher air stability of TTP-Br-capped NCs compared to that of DDA-Br-capped NCs, along with the fact that TTP-Br ligands are less bulky and, therefore, likely less electrically resistive than DDA-Br, motivated us to fabricate green-emitting LED devices based on TTP-Br-capped NCs. Our LEDs achieved a maximum EQE of 17.2 % and luminance of 2600 cd m$^{-2}$, surpassing the values reported to date for DDA-Br-based LED devices (~15%). Our work demonstrates that alkylphosphonium salts represent a promising class of ligands for the surface passivation of perovskite NCs. Further engineering of the groups bound to the phosphonium head could potentially lead to optimized ligands capable of boosting the performance of perovskite-based optoelectronic devices.

## ASSOCIATED CONTENT

**Supporting Information**

Additional data, including synthesis scheme, TEM and STEM characterization, XRD analysis, XPS, SEM elemental mapping (EDS), elemental analysis, NMR and FTIR characterizations, optical characterizations, UPS and DFT analyses, are presented.

## ACKNOWLEDGEMENTS

M.P., J.Z., P.R., L.D., and L.M. acknowledge funding from the Project IEMAP (Italian Energy Materials Acceleration Platform) within the Italian Research Program ENEA-MASE (Ministero dell'Ambiente e della Sicurezza Energetica) 2021-2024 "Mission Innovation" (agreement 21A033302 GU n. 133/5-6-2021). H. R. and F. S. acknowledge support by the European Research Council via the ERC-StG "NANOLED" (Grant 851794). The authors are thankful to Simone Lauciello for the SEM-EDS compositional analysis and Dorwal Marchelli for helping with the low-temperature PL measurements.

# Supporting Information for:

# Improving the Stability of Colloidal CsPbBr$_3$ Nanocrystals with an Alkylphosphonium Bromide as Surface Ligand Pair

Meenakshi Pegu[1,#], Hossein Roshan[2,#], Clara Otero-Martínez[1], Luca Goldoni[3], Juliette Zito[1], Nikolaos Livakas[1,4], Pascal Rusch[1], Francesco De Boni[3], Francesco Di Stasio[2], Ivan Infante,[5,6] Luca De Trizio[7*], Liberato Manna[1]*

[1] Nanochemistry, [2] Photonic Nanomaterials, [3] Materials Characterization, [7] Chemistry Facility, Istituto Italiano di Tecnologia, Via Morego 30, 16163 Genova, Italy

[4] Dipartimento di Chimica e Chimica Industriale, Università di Genova,16146 Genova, Italy

[5] BCMaterials, Basque Center for Materials, Applications, and Nanostructures, UPV/EHU Science Park, Leioa 48940, Spain

[6] Ikerbasque Basque Foundation for Science, Bilbao 48009, Spain

**Table S1.** List of phosphorous-based ligands used for perovskite NCs synthesis.

| Ligand | Perovskite NCs | Year | Efficiency | References |
|---|---|---|---|---|
| Trioctylphosphine | $CsPbBr_3$ | 2018 | PLQY = 50 %<br>EQE = 0.014 % | *Joule* 2.10 (2018): 2105-2116. |
| Ethane-1,2-diylbis(triphenyl phosphonium) bromide | $CsPbBr_3$ | 2019 | PLQE = 78 %<br>EQE = 6.3 % | *The Journal of Physical Chemistry Letters* 10.19 (2019): 5923-5928. |
| Triphenyl(9-phenyl-9H-carbazol-3-yl) phosphonium bromide | $CsPbBr_3$ | 2019 | PLQE = 77.6 % | *The Journal of Physical Chemistry Letters* 10.19 (2019): 5836-5840. |
| Phenylphosphonic dichloride | $CsPbCl_3$ | 2021 | PLQY = 71 %<br>EQE = 0.18 % | *ACS Energy Letters* 6.10 (2021): 3545-3554. |
| Trihexyl(tetradecyl)phosphonium bromide | $CsPbBr_3$ | 2023 | PLQY = 55 % | *The Journal of Chemical Physics* 158.17 (2023). |
| Triphenyl (9-phenyl-9H-carbazol-3-yl) phosphonium bromide | $CsPbBr_3$ | 2023 | EQE = 4.15 % | *ACS Energy Letters* 8.10 (2023): 4259-4266. |

**Experimental Section**

**Materials.**

Cesium carbonate (Cs$_2$CO$_3$, reagent plus, 99 %), lead acetate trihydrate (PbAc$_2$. 3H$_2$O, 99.99 %), oleic acid (OA, 90 %), benzoyl bromide (C$_6$H$_5$OBr, 97 %), octadecene (ODE, technical grade, 90%), toluene (anhydrous, 99.5 %), ethyl acetate (anhydrous, 99.8 %), trimethyl phosphine (PMe$_3$, 97%), 1-bromotetradecane (97 %), diethyl ether (HPLC grade, 99.9%) dimethyl sulfone (DMS, TraceCERT Certified Reference Materials for quantitative NMR, 99.99%), toluene d8 (99 atom % D), chloroform-d (CDCl$_3$, 99.8 atom % D), 1,4-dioxane (anhydrous, 99.5 %), chlorobenzene (anhydrous, 99.5 %), dimethylsufoxide-d6 (DMSO-d6, 99.9 atom % D), lithium fluoride (LiF, 99.9%), and Polyvinyl Carbazole (PVK) are purchased from Sigma Aldrich. Didodecylamine (DDAm, 97 %) is purchased from TCI. Poly(3,4-ethylenedioxythiophene):poly(styrene sulfonate) (PEDOT:PSS) with ratio of 1:6 (AI 4083) was purchased from Ossila. Patterned indium tin oxide (ITO) substrates, Poly(triaryl amine) (PTAA), and 2′,2′-(1,3,5-benzinetriyl)-tris(1-phenyl-1-H-benzimidazole) (TPBi, 98%) purchased from Lumtech. Aluminum pallets were purchased from Ted Pella for evaporation use. Unless otherwise stated, all the materials are used without further purification.

**General procedure for the synthesis.**

**Synthesis of trimethyl(tetradecyl)phosphonium bromide or TTP-Br ligand.**

To an oven-dried three-neck Schlenk round bottom flask equipped with a magnetic stir bar 1-Bromotetradecane (5.5 mmol) in anhydrous toluene (10 mL) is added and degassed for 5 minutes at room temperature. 5 mL of trimethyl phosphine (5 mmol) is added slowly under a nitrogen atmosphere, after which the mixture is slowly heated to 80°C and then stirred for 18 hours under reflux conditions. After reaction completion, the mixture is concentrated under reduced pressure to obtain a crude solid. This solid is further washed with diethyl ether (5 times x 20 mL) using a Buchner funnel and dried under vacuum at room temperature, to obtain the final product as TTP-Br. White solid obtained 1.32 g, yield 75 %. The chemical structure and purity of the TTP-Br ligand were ascertained by $^1$H NMR and $^{13}$C NMR in toluene d8 (see Supporting Information for peak assignment, **Figures S1-S3**).

**Scheme S1.** Synthesis route for trimethyl(tetradecyl)phosphonium bromide (TTP-Br).

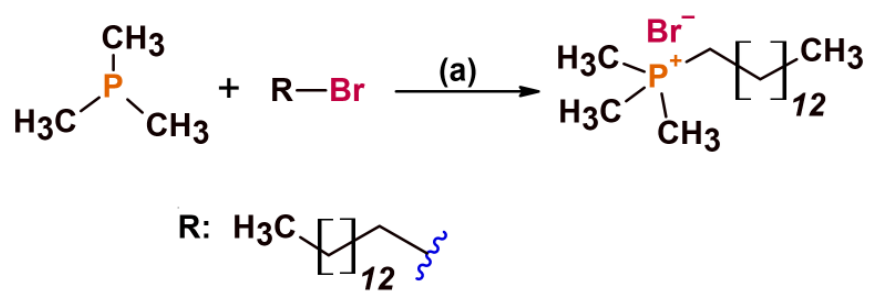

Reaction condition: (a) 10 mL toluene, 80°C, 18h reflux.

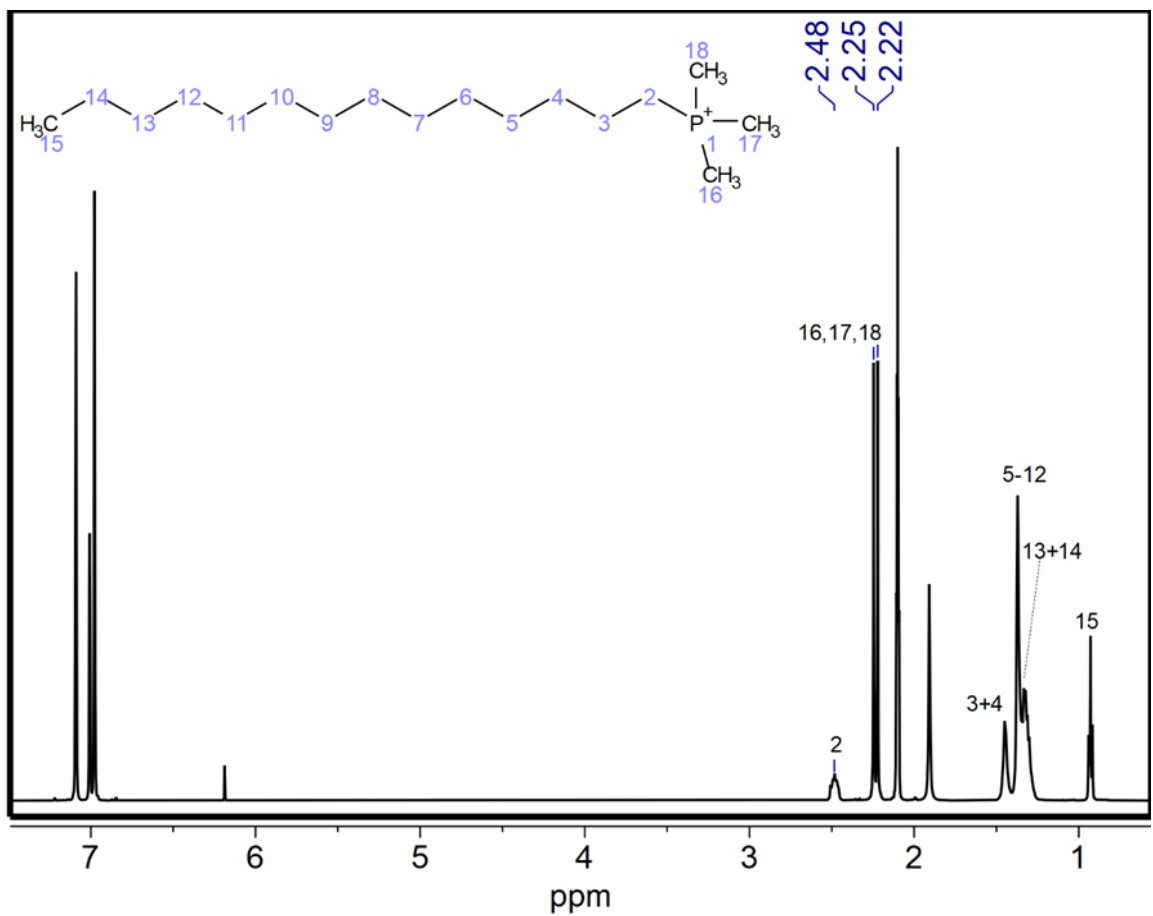

**Figure S1.** $^1$H NMR spectrum of TTP-Br in toluene-D at 298 K, with peak assignment, the TTP-Br structure formula, and signal numbering.

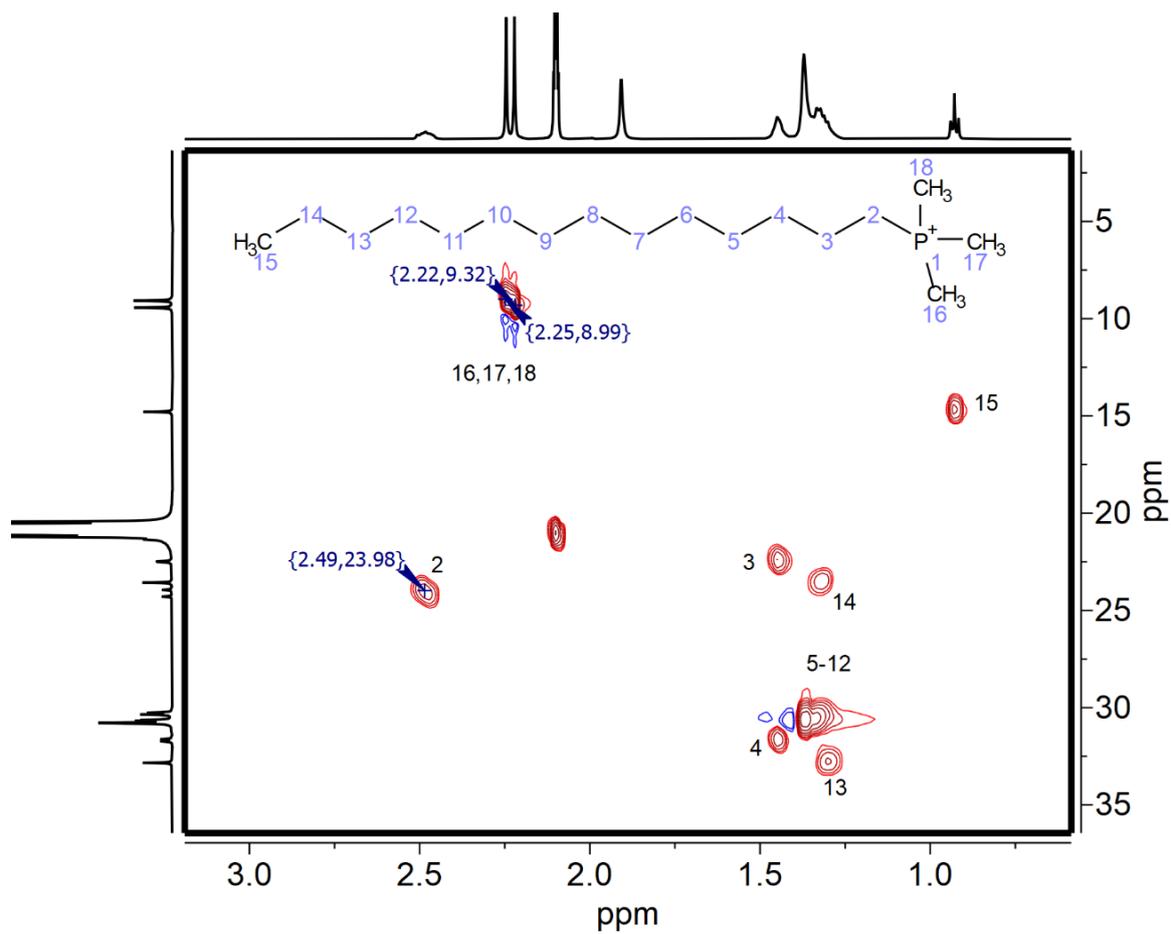

**Figure S2.** $^1$H-$^{13}$C HSQC NMR spectrum of TTP-Br in toluene-D at 298K, with peak assignment, the TTP-Br structure formula, and signal numbering.

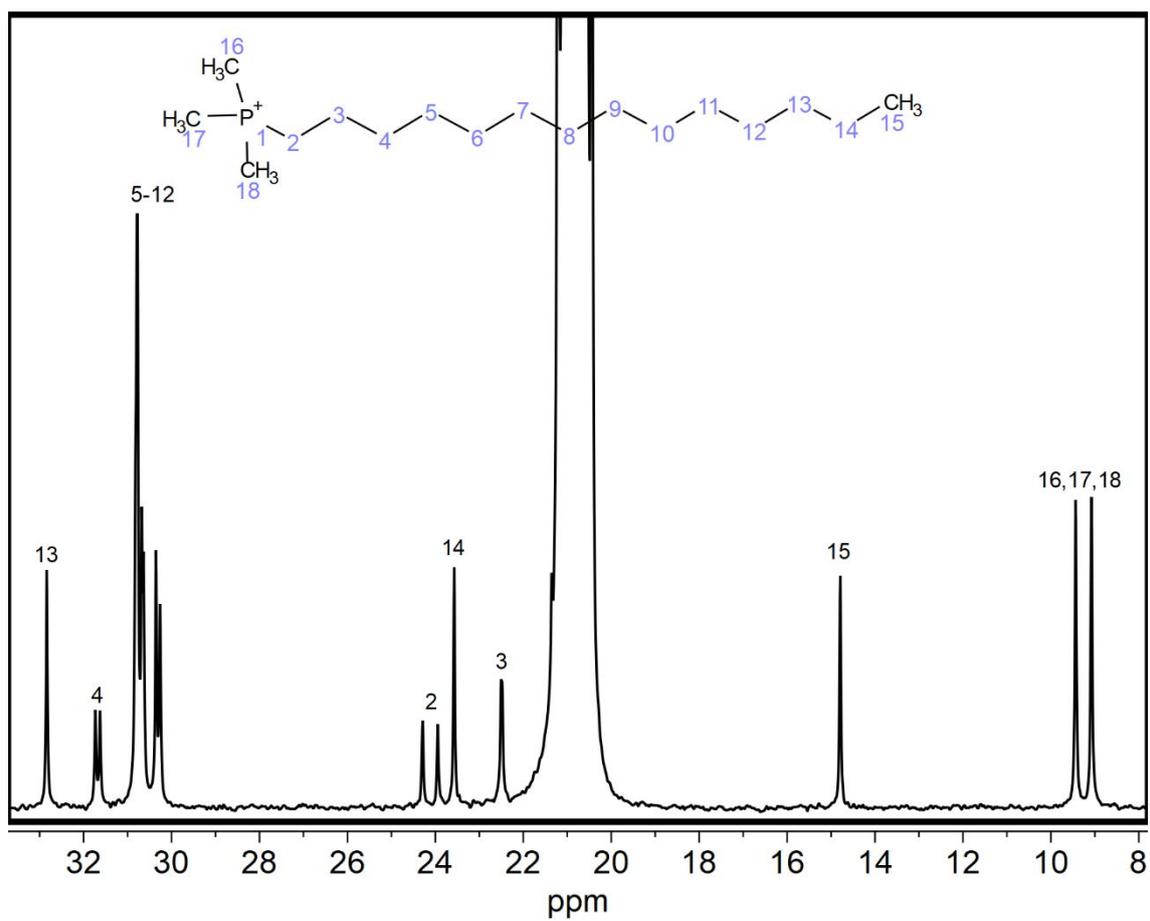

**Figure S3.** $^{13}$C {$^{1}$H} decoupled NMR spectrum of TTP-Br in toluene-D at 298K, with peak assignment, the TTP-Br structure formula, and signal numbering.

**Preparation of Cs-Pb-Oleate stock solution.**

Pb(CH$_3$COO)$_2$·3 H$_2$O (760 mg), Cs$_2$CO$_3$ (160 mg), and oleic acid (15 mL) are loaded in a 25 mL three-neck flask. The mixture is heated up to 100°C and degassed using a Schlenk line for 1.5 hours. At this point, the precursor salt is completely dissolved, and the solution is cooled down to room temperature and stored in a glove box.

**Preparation of Didodecylamine stock solution.**

In a glovebox, 10 mL of anhydrous toluene is added to 4.43 g of Didodecylamine (DDA) in a 25 mL vial. The mixture is then placed on a hot plate and heated to 60°C for 10 minutes until the DDA in the mixture is completely dissolved. After cooling, the mixture is used for the subsequent synthesis.

**Synthesis of CsPbBr$_3$ NCs.**

The Cs-Oleate-capped CsPbBr$_3$ NCs are synthesized following the protocol of Imran et al. reported using the standard Schlenk technique.[1, 2] Firstly, 1.5 mL of Cs-Pb-Oleate stock solution, 1.5 mL of Didodecylamine (DDA) stock solution, and 9 mL of 1-octadecene are loaded together into a 40 mL glass vial. The mixture is heated up to 90°C and degassed for 2 hours. It is then heated up to 100°C under nitrogen, and at that point, a solution of benzoyl bromide (50 μL) in anhydrous toluene (500 μL) is injected swiftly. After 60 seconds, the reaction is quenched by cooling in an ice-water bath and directly used for ligand exchange. The crude Cs-Oleate-capped CsPbBr$_3$ NCs solution is divided into three different fractions (3 mL each); one fraction is used as the reference, and the other two fractions are used for post-treatment ligand exchange. The organic solvents used in the present work were anhydrous to avoid moisture contamination. All the samples have been cleaned to remove the excess ligands prior to their structural and optical characterizations.

**Ligand exchange and washing of NCs.**

Each of the three fractions (3 mL) of crude Cs-Oleate-capped CsPbBr$_3$ NCs solution is used for the post-synthesis ligand exchange treatment with 2 mL DDA-Br (25 mM), and 2 mL TTP-Br (25 mM), respectively. TTP-Br halide salts have lower solubility in nonpolar solvents such as toluene. In this case, we added a small amount of chloroform (0.5 mL) to dissolve TTP-Br in toluene (4.5 mL) and heated at 80ºC for five minutes. The complete solubility of TTP-Br maximized the passivating effect on the perovskite NCs' surface. A similar exchange protocol was followed for the DDA-Br-capped NCs. The vials containing the mixtures of DDA-Br-capped and TTP-Br-capped NCs are then stirred for 20 mins under a nitrogen atmosphere. After that, the NCs are precipitated and washed, adding an excess of anhydrous ethyl acetate (20 mL), and separated upon centrifugation at 6000 rpm. The supernatant is discarded to remove the unreacted residues and ligands from the precipitate and is redispersed in anhydrous toluene (1 mL). The washing process was repeated by adding ethyl acetate (6 mL) for the second time, centrifuged at 6000 rpm, and finally redispersed in neat toluene (1mL).

The overall washing procedure was carried out two times following the same procedure with pure anhydrous toluene only (without using ligands) for the crude Cs-Oleate-capped CsPbBr$_3$ NCs. Organic solvents used in the present work are anhydrous to avoid moisture contamination.

We have observed that further addition of TTP-Br ligand after the second wash resulted in the degradation of the NCs, adversely affecting both their size and morphology (**Figure S42**).

**LED Fabrication using TTP-Br-capped NCs.**

First, the patterned ITO glass substrate was ultrasonically cleaned sequentially using an ITO cleaning solution, acetone, and isopropanol, then dried with a nitrogen flow. Subsequently, substrates underwent ultraviolet-ozone treatment to achieve a hydrophilic surface. For the LED devices, a PEDOT:PSS solution was spin-coated onto the treated ITO glass at 5000 rpm for 40 seconds and then annealed at 120°C for 10 minutes under ambient conditions. The ITO/PEDOT:PSS substrates were then transferred to a glovebox where a single-hole transport layer of PTAA (10 mg/mL in chlorobenzene) or PVK (3 mg/mL in 1,4-dioxane) was spin-coated at 2000 rpm for 45 seconds and annealed at 75°C for 10 minutes. For a double-hole transport layer, PTAA (10 mg/mL) was first spin-coated and annealed at 75°C for 10 minutes. Once cooled, an upper PVK layer (3 mg/mL) was applied using dynamic spin coating. The emitting layer (TTP-Br-capped $CsPbBr_3$ NCs) was formed by dynamic spin-coating of the NCs solution at 2000 rpm for 45 seconds. Electron transport layers of TPBi (35 nm), LiF (1 nm), and an aluminum electrode (100 nm) were sequentially deposited in a vacuum chamber through a shadow mask at a base pressure of $2 \times 10^{-6}$ Torr. The thickness of the deposited layers was monitored using a quartz crystal. The active device area, defined by the overlap between the ITO and aluminum electrodes, was approximately 4 mm².

**LED Characterization.**

The JVL properties were assessed utilizing a Keithley 2636 source-measure unit in conjunction with a Thorlabs PDA 100 A silicon switchable gain detector. The silicon detector's output was transformed into power (photon flux) based on its responsivity. All measurements were done in the air with almost 50% of relative humidity. For luminance measurements, the surface of ITO/glass was considered as the light output without using an integrated sphere. To determine the EQE, we calculated the ratio of photon flux to the device's driving current. EL spectra were gathered using an Edinburgh Instrument FLS 900 spectrometer step scan.

**Methods.**

**Powder X-ray diffraction (XRD) analysis.**

XRD patterns are acquired using the PANanalytical Empyrean X-ray diffractometer equipped with a 1.8 kW Cu Kα ceramic X-ray tube and a PIXcel3D 2x2 area detector, operating at 40 mA and 45 kV using parallel beam geometry and symmetric reflection mode. The NC samples for diffraction measurements are prepared by concentrating the NC solution under nitrogen and drop-casting the dispersions on a zero-diffraction Si substrate. XRD patterns were acquired at room temperature under ambient conditions.

**Transmission Electron Microscopy (TEM).**

Conventional bright-field TEM images are acquired on samples prepared by drop casting the diluted NC colloids onto 200 mesh carbon film-coated copper grids using a JEOL-JEM1400Plus transmission electron microscope operating at an acceleration voltage of 120 kV. Scanning TEM coupled with energy-dispersive X-ray spectroscopy (EDS) was performed on samples prepared in the same way and which were stored under dynamic vacuum conditions for 24 hours, using an image-aberration corrected JEOL JEM2200FS microscope with Schottky emitter operated at 200 kV and a Bruker X-Flash 5060 silicon-drift detector with 60 mm$^2$ area.

**Optical Absorption and Photoluminescence Spectroscopy.**

Optical absorption and photoluminescence spectra were recorded using a Varian Cary 300 UV-Vis spectrophotometer and a Varian Cary Eclipse spectrofluorometer using an excitation wavelength of 350 nm for all the samples. The samples are prepared by diluting NC colloids dispersed in toluene in quartz cuvettes with a 1 cm path length.

**Photoluminescence Quantum Yield (PLQY) and Time-Correlated Single-Photon Counting (TCSPC).**

The PLQY of the NC samples is measured using an FS5 Spectrofluorometer-Edinburgh Instruments, equipped with the integrating sphere with step increments of 0.5 nm and an integration time of 0.2 s per data point for one scan. The NC samples are prepared by diluting 10-15 μL of the concentrated NC solutions in 3 mL of anhydrous toluene within quartz cuvettes (1 cm path length), capped with white PTFE stoppers (Helma-Analytics, part number 111-10-40). A blank reference sample is prepared using 3 mL of anhydrous toluene. All the NC solutions are then diluted to obtain an optical density of ~ 0.20 - 0.21 at 375 nm excitation wavelength and measured with the integrating sphere using the output of the continuous Xenon lamp.

Time-resolved photoluminescence spectra (TRPL) at room temperature are obtained using an FLS900 Edinburgh spectrophotometer, and the PL decay traces are measured with a pulsed laser diode (λex = 375 nm, 10 MHz repetition rate, 60 ps pulse width) and fitted with a three-exponential decay function.

**X-ray Photoluminescence Spectroscopy (XPS).**

X-ray photoelectron spectroscopy (XPS) measurements were carried out through a Kratos Axis Ultra$^{DLD}$ spectrometer (Kratos Analytical Ltd.) with a monochromated Al K$_α$ X-ray source (*hv* = 1486.6 eV) operating at 20 mA and 15 kV. Specimens were prepared by dropping a concentrated NCs solution in toluene onto a highly ordered pyrolytic graphite (HOPG, ZYA grade) substrate. The wide scans were collected over an analysis area of 300 × 700 µm² at a photoelectron pass energy of 160 eV and energy step of 1 eV, while high-resolution spectra of Cs 3d, Pb 4f, Br 3d, and C 1s, are collected at a photoelectron pass energy of 20 eV and an energy step of 0.1 eV. A take-off angle of 0° concerning the sample's normal direction was used for all analyses. The slight differential electrical charging effects (less than 0.5 eV) observed on all samples were not neutralized. The irradiation of an APbX$_3$-like perovskite with an electron beam is known to induce the desorption of halogen species and the nucleation of metallic Pb particles.[3] The spectra have been referenced to the adventitious carbon 1s peak at 284.8 eV. The spectra were analyzed with the Casa XPS software (Casa Software Ltd., version 2.3.24),[4] and the residual background was eliminated by the Shirley method across the binding energy range of the peaks of interest. The relative atomic concentrations were then estimated using the specific function in the Casa XPS software.

**Ultraviolet photoelectron spectroscopy (UPS)**

The ultraviolet photoelectron spectroscopy (UPS) measurements were performed using a He I (21.22 eV) discharge lamp, fitted in the same chamber used for XPS analyses, on an area of 55 µm in diameter, at a pass energy of 10 eV and with a dwell time of 100 ms. The energy levels within the equipment and UPS-based calculations are described as follows. The work function, φ, i.e. the position of the Fermi level versus the vacuum level, was determined for each sample from the position of the secondary electron cutoff in the UPS spectrum, using the following equation: **φ = hν – E$_0$**, where hν is the source energy (21.22 eV for He I photons) and E$_0$ is the secondary electron cut off. The position of the valence band maximum (VBM) versus the vacuum level, i.e. the ionization energy, E$_{ion}$, was determined from the width of the entire UPS spectrum, according to the following equation: **E$_{ion}$ = hν – (E$_0$ – E$_1$) = hν – E$_0$ + E$_1$**, where E$_1$ is the position of the VBM concerning the zero (Fermi) level.[5] The values E$_0$ and E$_1$ were determined in the UPS spectrum through the background functions "Edge Up" and "Edge Down", respectively, in the Casa XPS software. The error bar associated with this procedure was estimated to be equal to 0.2 eV.

**Nuclear magnetic resonance (NMR).**

NMR was performed on a Bruker Avance III 600 MHz (600.13 MHz) spectrometer, fit with a 5 mm QCI cryoprobe. The matching, tuning, and line shape resolution were adjusted and the 90° pulse was finally calibrated by an automatic pulse calculation routine, before acquisition.[6] The temperature was actively monitored on each sample tube and the sample was let to equilibrate 2 minutes inside the probe earlier than the pre-acquisition routines.

*$^1$H NMR spectra* were acquired at 298 K and 313 K, by accumulating 32 scans (64 for *q*-NMR and NCs) without steady ones, with inter-pulses delay of 30s (64 for *q*-NMR and NCs) and 65536 digit points, over a spectral width of 20.83 ppm with the offset positioned at 6.18 ppm.

Spectra were smoothed with an exponential function equivalent to 0.3 Hz before Fourier transform. The ligand concentration was determined in DMSO-D6 (Deuterated Dimethyl sulfoxide) by comparing the ligand signal integrated intensity with that of a 10 mM DMS standard solution freshly prepared, by using the PULCON (Pulse Length Based Concentration Determination) method.[7]

The $^1$H–$^{13}$C HSQC (Heteronuclear Single Quantum Coherence) experiments were performed with 16 transients (64 for NCs), 2048 digit points, 256 increments and $^1$J CH of 145 Hz, a spectral width of 15.15 ppm for $^1$H and 165.8 ppm for $^{13}$C, with a transmitter frequency offset positioned at 7.00 and 75.0 ppm, respectively.

The $^1$H-$^1$H NOESY (Nuclear Overhauser Spectroscopy) experiments were acquired with 32 scans (64 for NCs and for NOESY at 313K), a mixing time of 300 ms, over a spectral width of 15.15 ppm, centered at 7.49 ppm.

$^{13}$C NMR spectra were acquired by using a 30° for the $^{13}$C pulse excitation and a broadband decoupling for $^1$H, with 18435 transients, a relaxation delay of 2s, over a spectral width of 236.65 ppm centered at 100.00 ppm.

**Surface ligand density calculations**.

The surface ligand density of the samples was estimated by calculating the ratio between the concentration of ligands to the NCs' surface obtained by NMR and ICP-OES, respectively. The concentrations of both ligand and Pb were obtained after evaporating the colloidal solution of the NCs in deuterated toluene (previously characterized). The resulting precipitate was thoroughly dried under a nitrogen flow and subsequently dissolved in the dried NCs in deuterated-DMSO (200 μL). The NCs in the DMSO-d6 solution were then loaded into a SampleJet NMR tube of 3 mm for ligand quantification.

***Ligand quantification.*** The concentration of different ligands (TTP-Br, DDA-Br, and Cs-Oleate) was quantified using *quantitative* NMR spectroscopy by comparing the integrated area of ligand signals, each normalized for the number of protons generating the peak (2H), to that of a solution of dimethyl sulfone 10 mM (TraceCERT®) used as a standard solution, in DMSO-$d_6$, the latter normalized to 6 H. The concentration of all the ligands was obtained by the integrated peak ratio between a proton signal which is the characteristic peak of each ligand, and that of the external standard solution through the PULCON (PUlse Length-based CONcentration determination) method[7]. In the case of TTP-Br, the signal employed was the multiplet at 2.14 ppm. For Cs-Oleate-capped NCs, the signal at 5.30 ppm which corresponds to the protons of the alkyl double bond of the oleate species was measured. Finally, the triplet at 3.20 ppm corresponding to the methyl group of the DDA-Br ligand was used to estimate its concentration.

***Inductively Coupled Plasma−Optical Emission Spectroscopy (ICP-OES).*** After the ligand quantification on the deuterated DMSO solution, the Pb concentration of the NCs was estimated by ICP-OES on an aiCAP 6000 spectrometer (Thermo Scientific) Prior to the measurement, 50 µL of the DMSO – $d_6$ sample were diluted to 25 mL in an aqueous solution of aqua regia (1:25) and subjected to an acid digestion overnight.

***Ligand density.*** The ligand density of the three samples was estimated by calculating the ratio of the ligand concentration and NC surface in the deuterated-DMSO solution:

$$Ligand\ density\ (ligand/nm^2) = \frac{[Ligand](mL^{-1})}{Total\ NC\ surface\ (nm^2/mL)}$$

The total NC surface is determined based on the Pb concentration in the DMSO-$d_6$ sample (measured by ICP analysis) and the size distribution of the NCs considering a unit cell size of 0.589 nm:

$$Total\ NC\ surface\ (nm^2/mL) = NC\ concentration\ (mL^{-1}) * [NC\ size\ (nm)]^2 * 6$$

$$NC\ concentration\ (mL^{-1}) = \frac{Total\ concetration\ of\ Pb\ (mL^{-1})}{Pb\ atoms/NC}$$

$$Pb\ atoms/NC = \left[\frac{NC\ size\ (nm)}{0.589\ nm/unit\ cell}\right]^3$$

**Density functional theory (DFT) calculations**.

We conducted atomistic simulations at the DFT level using the PBE exchange-correlation functional[8] and a double-ζ basis set with polarization functions[9], as implemented in CP2K version 6.1[10]. All structural optimizations were performed in a vacuum. Scalar relativistic effects were accounted for using effective core potential functions within the basis set. While spin-orbit coupling effects were not included, previous studies on similar systems have demonstrated that their influence on relaxed structural properties is negligible. Details regarding the model construction are provided in the main text as well as in the references[11, 12].

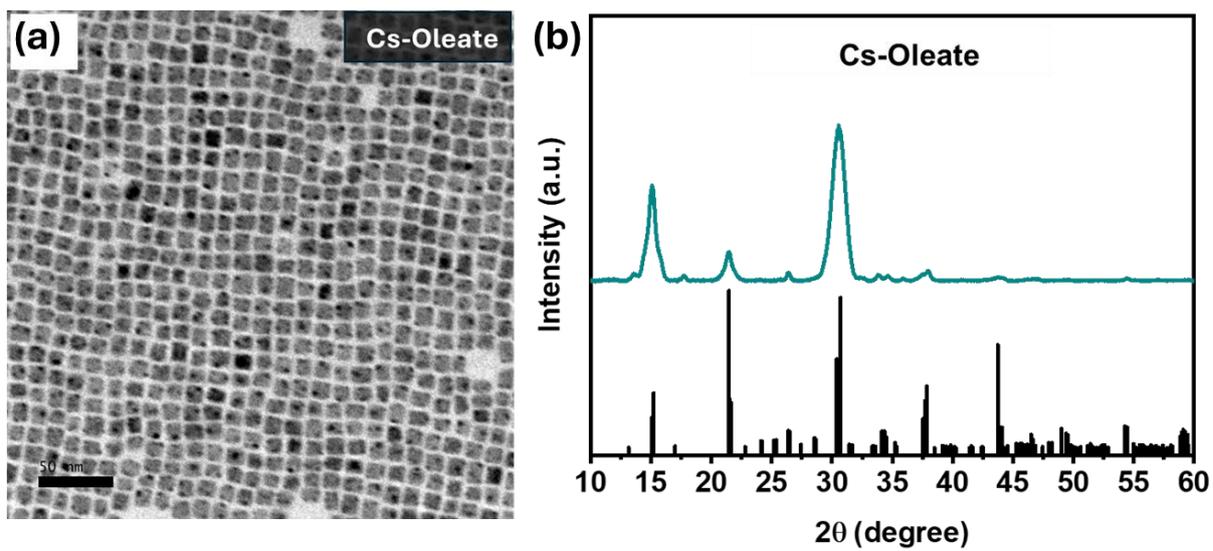

**Figure S4.** (a) TEM micrograph, and (b) XRD spectrum of Cs-Oleate-capped CsPbBr$_3$ NCs.

**Table S2.** Quantitative elemental analysis was obtained through XPS.

| CsPbBr$_3$ NCs | Cs (%) | Pb (%) | Br (%) | Cs/Pb | Br/Pb | N (%) | O (%) | P (%) |
|---|---|---|---|---|---|---|---|---|
| **Cs-Oleate** | 23.9 | 23.2 | 52.9 | 1.03 | 2.28 | 0 | 9.3 | 0 |
| **DDA-Br** | 22.5 | 20.5 | 57.0 | 1.10 | 2.78 | 2.7 | 2.5 | 0 |
| **TTP-Br** | 22.1 | 20.8 | 57.1 | 1.06 | 2.75 | 0 | 2.8 | 2.6 |

Elemental analysis using X-ray photoelectron spectroscopy (XPS) revealed the exchange of Cs-Oleate-capped CsPbBr$_3$ NCs with DDA-Br and TTP-Br ligands. A detailed explanation is provided in the manuscript. The observed reduction in the atomic percentage of Cs suggests the replacement of the Cs$^+$ with the PMe$_3^+$ group. Notably, N $1s$ peaks appear only in DDA-Br-capped CsPbBr$_3$ NCs. The increase in Br/Pb ratio indicates halogen compensation during the treatment of the quaternary ammonium and phosphonium salts. The relative atomic percentages of the elements (Cs, Pb, Br, O, N, C, P) are presented in **Table S2**.

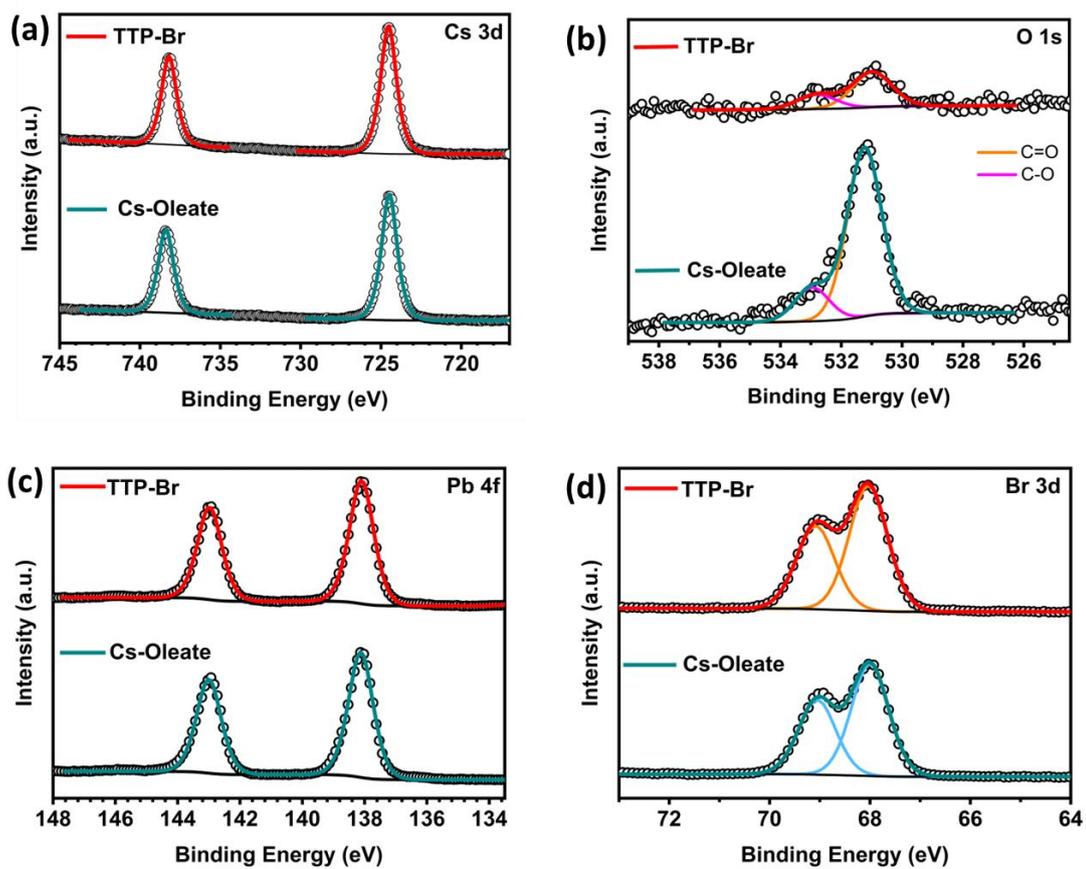

**Figure S5**. Comparison between (a) Cs 3*d*, (b) O 1*s*, (c) Pb 4*f*, (d) Br 3*d*, XPS spectra of TTP-Br-, and Cs-Oleate-capped CsPbBr$_3$ NCs.

.

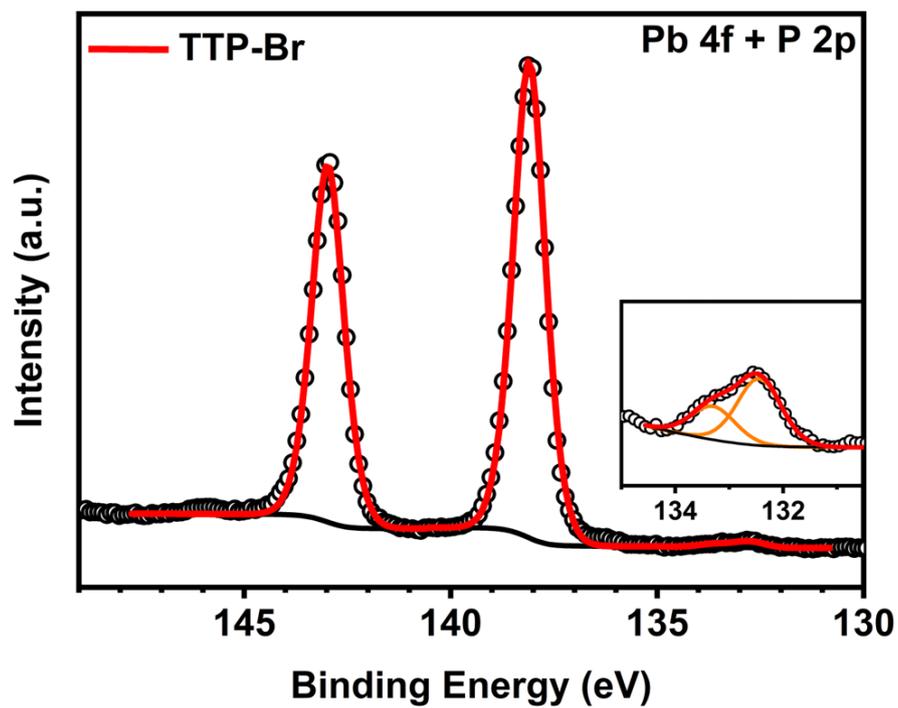

**Figure S6.** Pb 4*f* and P 2*p* (inset) XPS spectra of TTP-Br-capped CsPbBr$_3$ NCs.

**Table S3.** Elemental composition of the area shown in **Figure S7** performed via STEM-EDS.

| Map Sum Spectrum | Line Type | weight% | atomic% | rel. error in % (1 Sigma) |
|---|---|---|---|---|
| **P** | K series | 0.36 | 1.40 | 11.81 |
| **Br** | K series | 36.89 | 55.16 | 3.14 |
| **Cs** | L series | 22.53 | 20.25 | 10.15 |
| **Pb** | L series | 40.22 | 23.19 | 10.15 |
| **Total** | | 100.00 | 100.00 | |

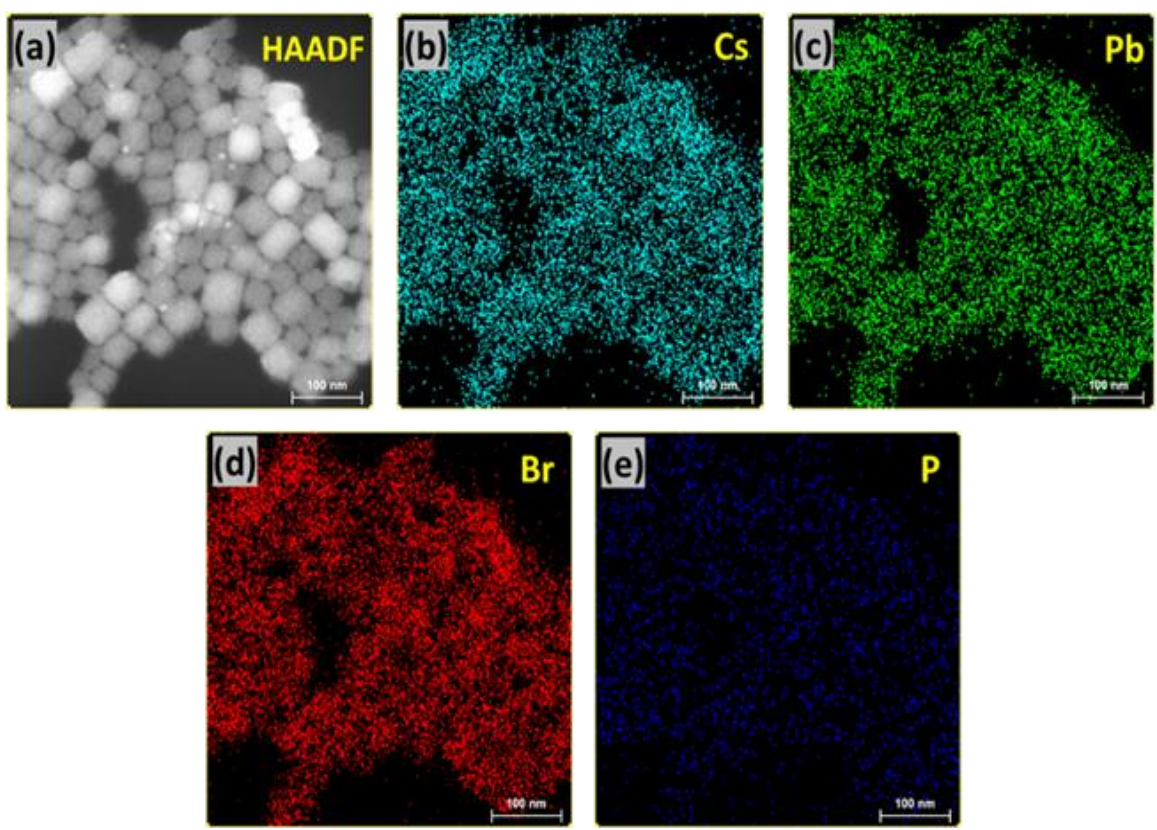

**Figure S7.** (a) Scanning transmission electron microscopy (STEM)-High-angle annular dark field (HAADF) image of TTP-Br-capped CsPbBr$_3$ NCs. (b-e) Corresponding elemental mappings of, Cs, Pb, Br, and P of TTP-Br-capped CsPbBr$_3$ NCs by energy-dispersive X-ray spectroscopy (EDS).

**Table S4.** Elemental analysis of the area shown in **Figure S8** derived from SEM-EDS.

| Map Sum Spectrum | Line Type | Wt%   | Wt% Sigma | Atomic % |
|------------------|-----------|-------|-----------|----------|
| **P**            | K series  | 1.34  | 0.03      | 4.90     |
| **Br**           | K series  | 39.14 | 0.12      | 55.56    |
| **Cs**           | L series  | 22.73 | 0.08      | 19.40    |
| **Pb**           | L series  | 36.80 | 0.13      | 20.15    |
| **Total**        |           | 100.00|           | 100.00   |

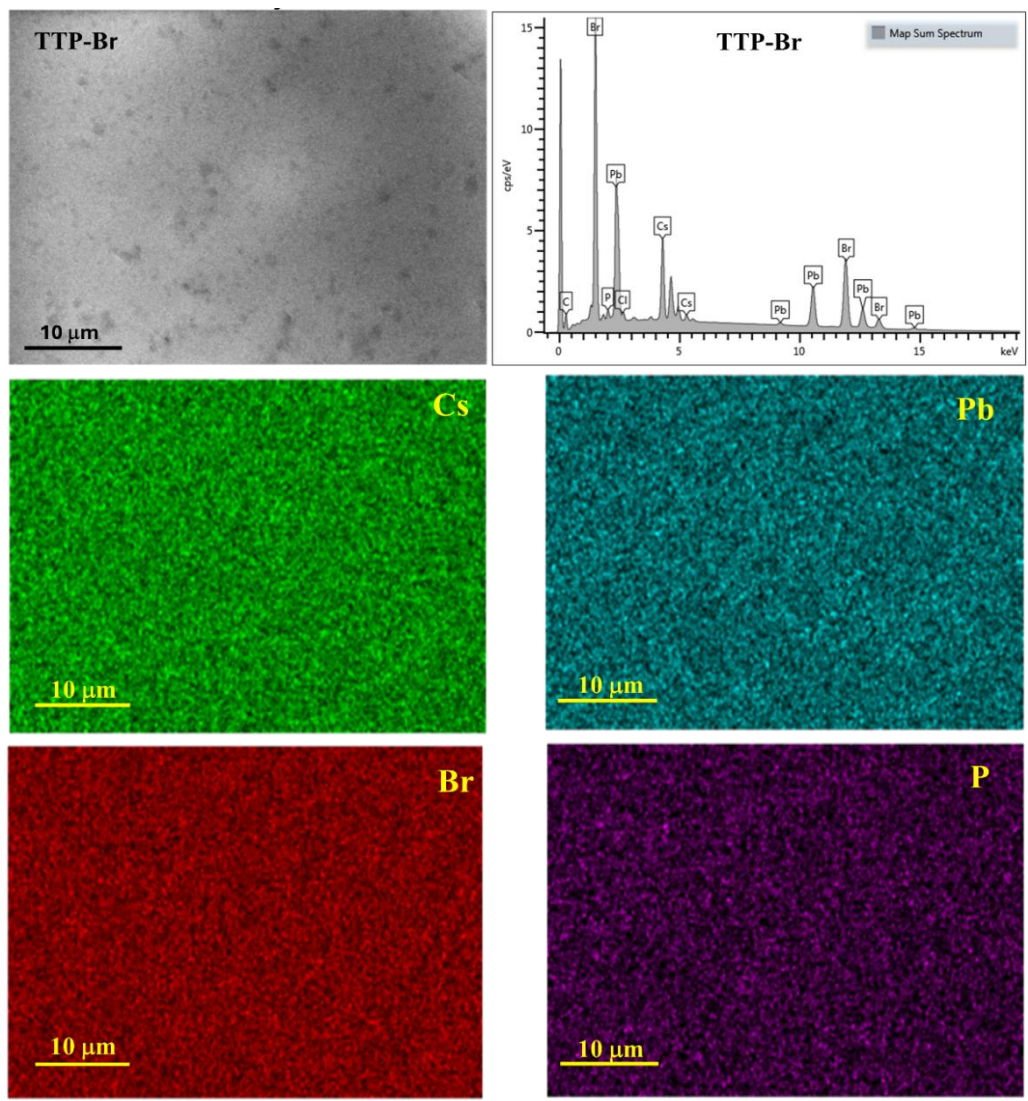

**Figure S8.** SEM-EDS elemental mapping of TTP-Br-capped CsPbBr$_3$ NCs.

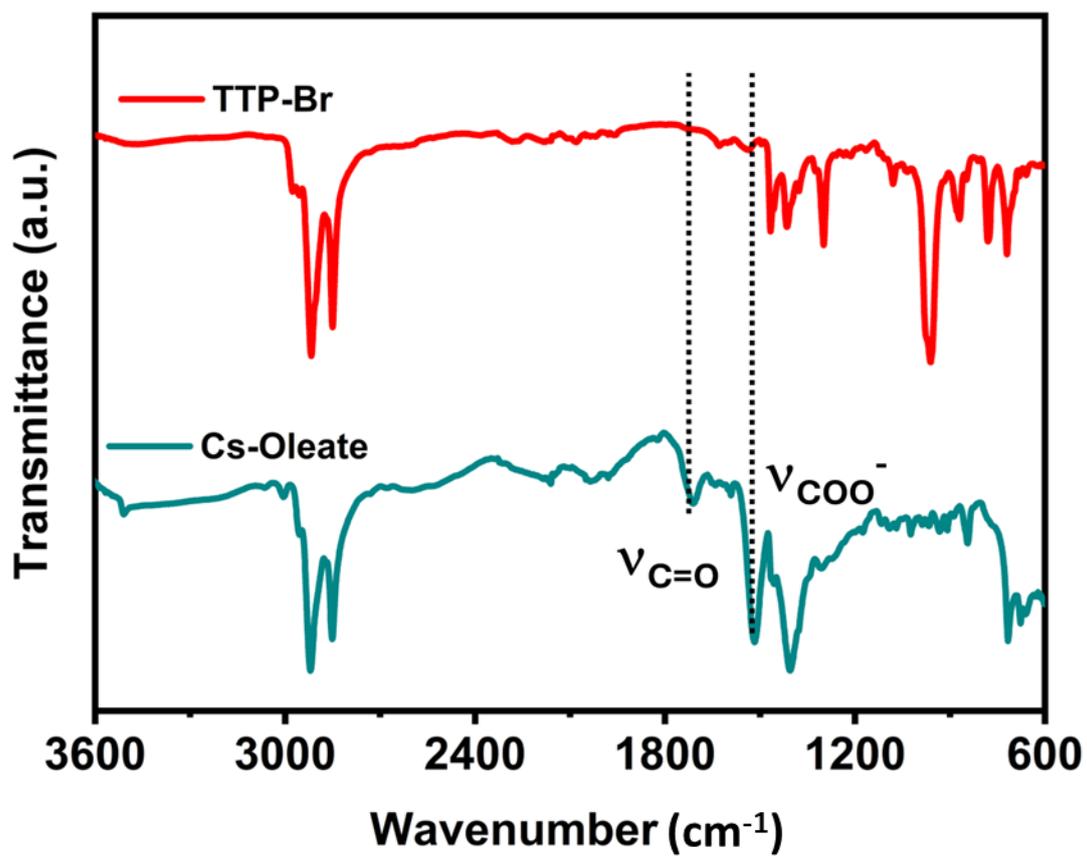

**Figure S9**. FT-IR spectra of TTP-Br-, and Cs-Oleate-capped CsPbBr$_3$ NCs.

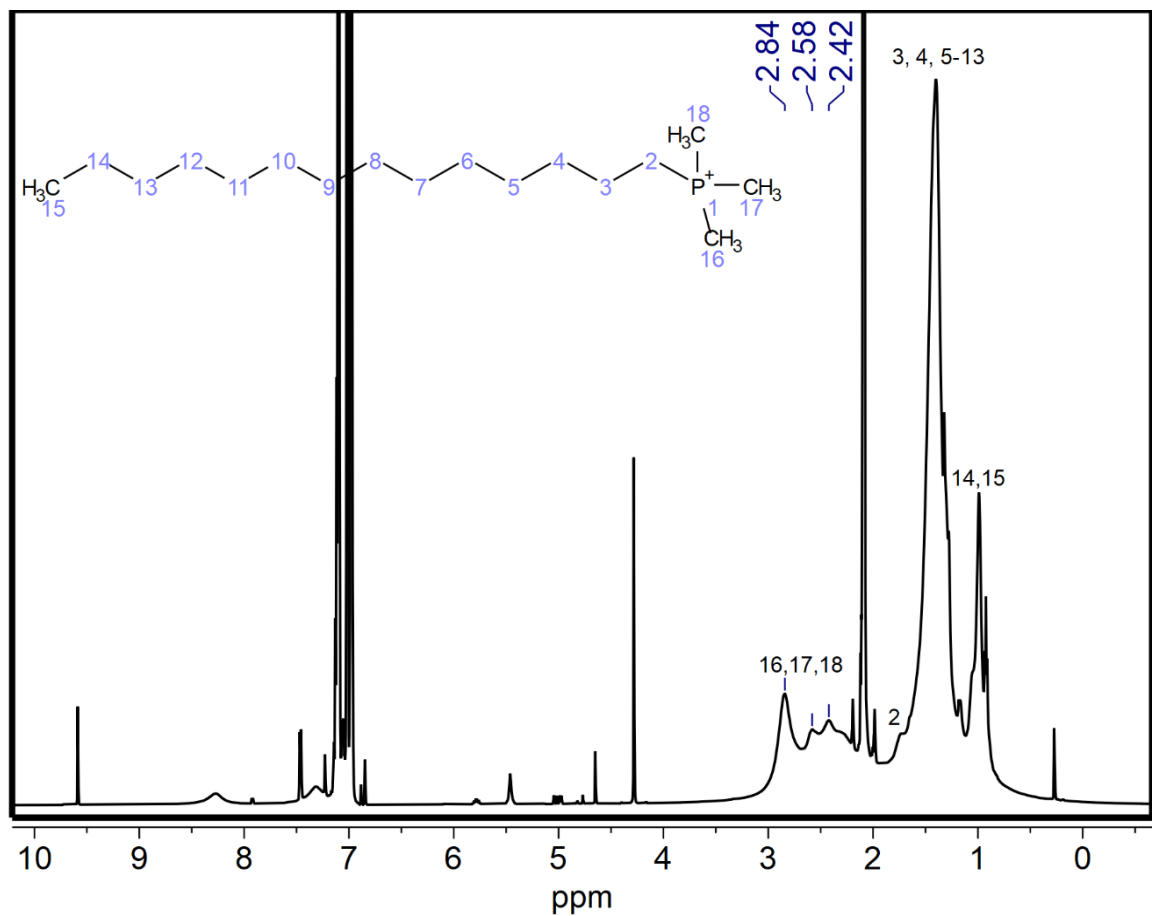

**Figure S10.** ¹H NMR spectrum of TTP-Br-capped CsPbBr₃ NCs in toluene-D at 298K, with diagnostic peak assignment, the TTP-Br structure formula, and signal numbering.

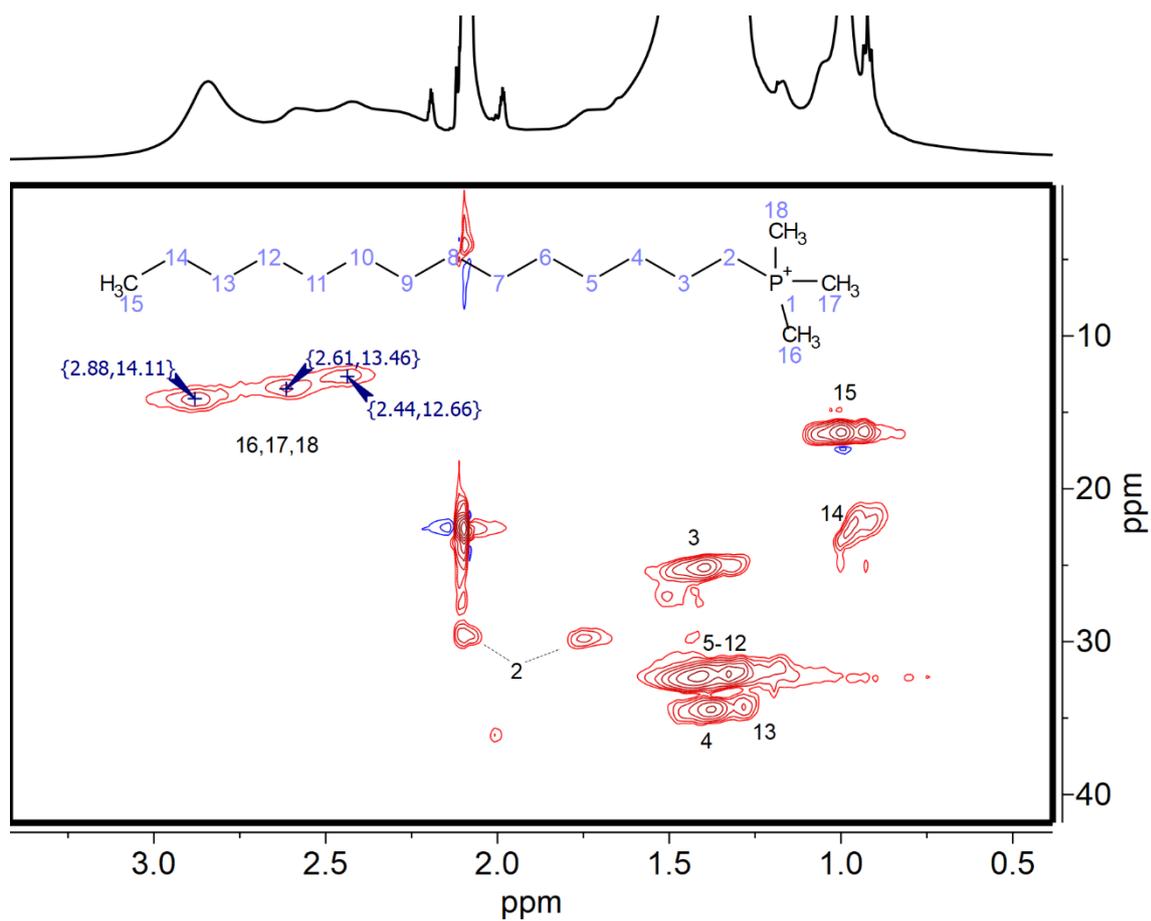

**Figure S11.** $^1$H-$^{13}$C HSQC NMR spectrum of TTP-Br-capped CsPbBr$_3$ NCs in toluene-D at 298K, with peak assignment, the TTP-Br structure formula, and signal numbering.

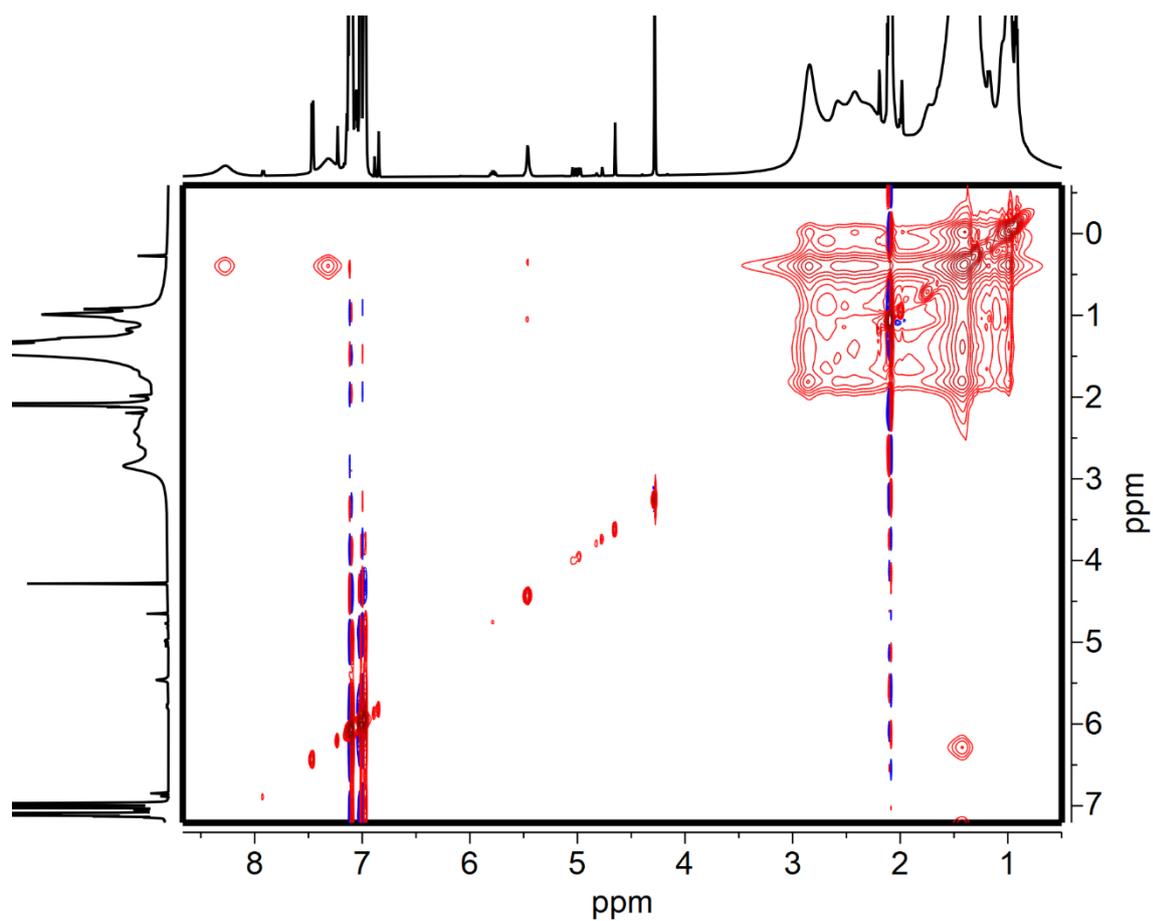

**Figure S12.** $^1H$-$^1H$ NOESY NMR spectrum of TTP-Br-capped CsPbBr$_3$ NCs in toluene-D at 298 K, negative (red) NOE cross peaks are typical of species with slow tumbling regime in solution, due to the binding with the NC's surface.

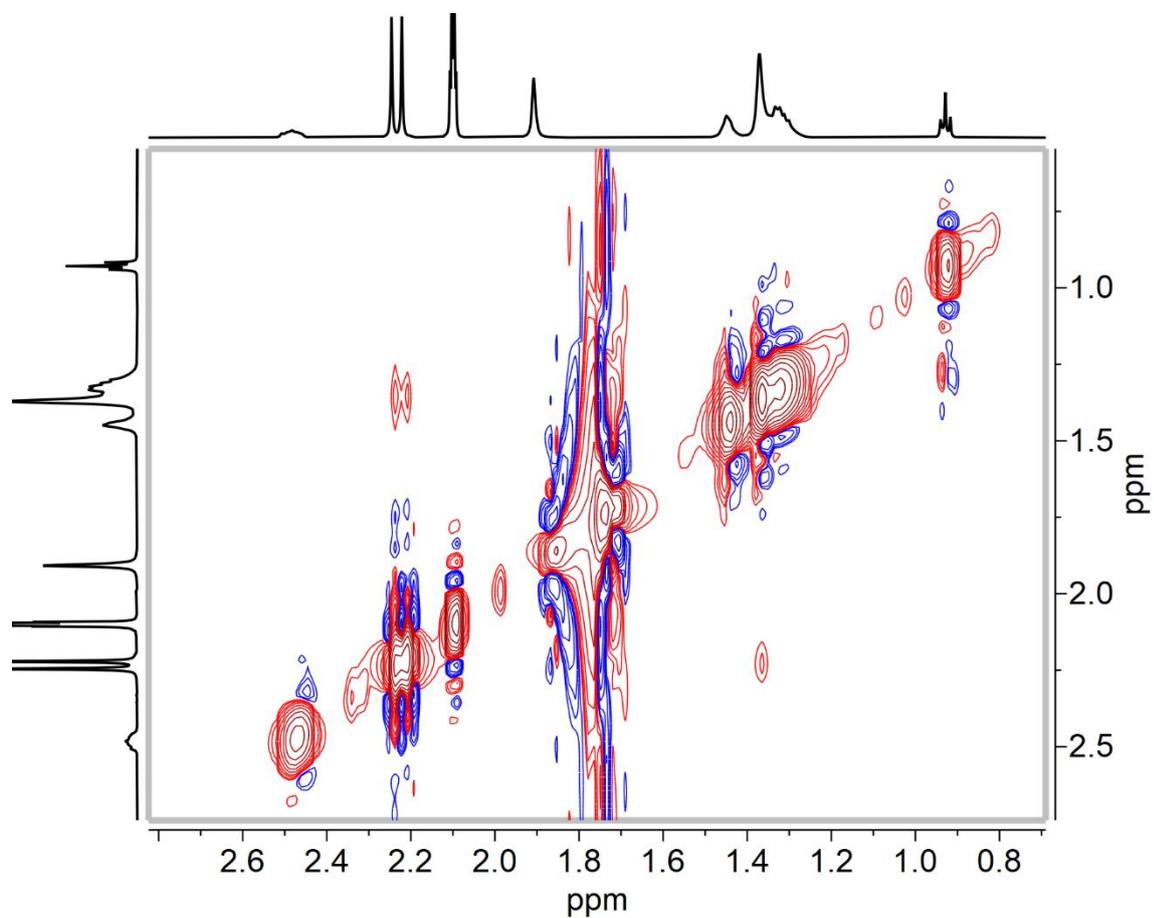

**Figure S13.** $^1H$-$^1H$ NOESY NMR spectrum of TTP-Br in toluene-D at 298 K, which shows negative (red) NOE cross peaks, likely due to micelles or aggregates formation. The same experiment at 313 K returns positive (blue) NOE cross peaks (**Figure S15**).

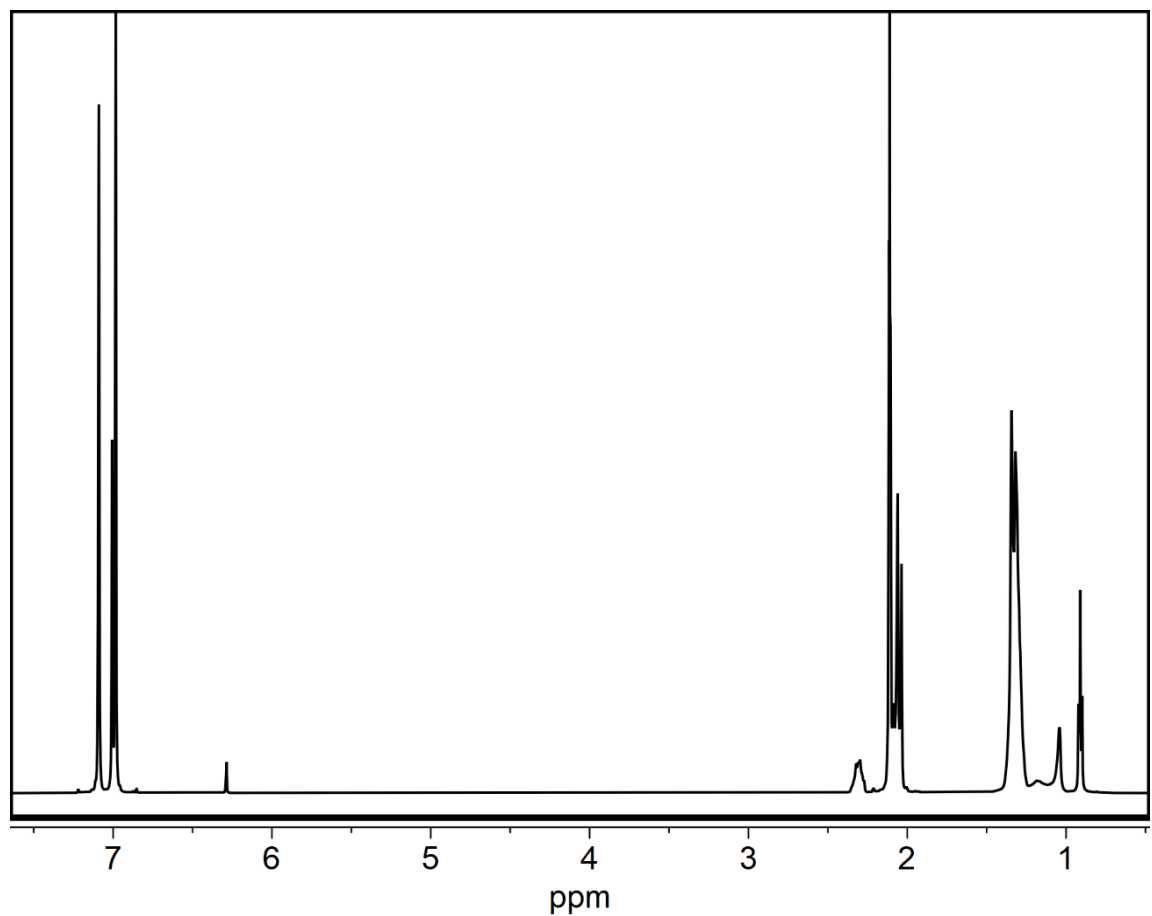

**Figure S14.** ¹H NMR spectrum of TTP-Br in toluene-D at 313 K.

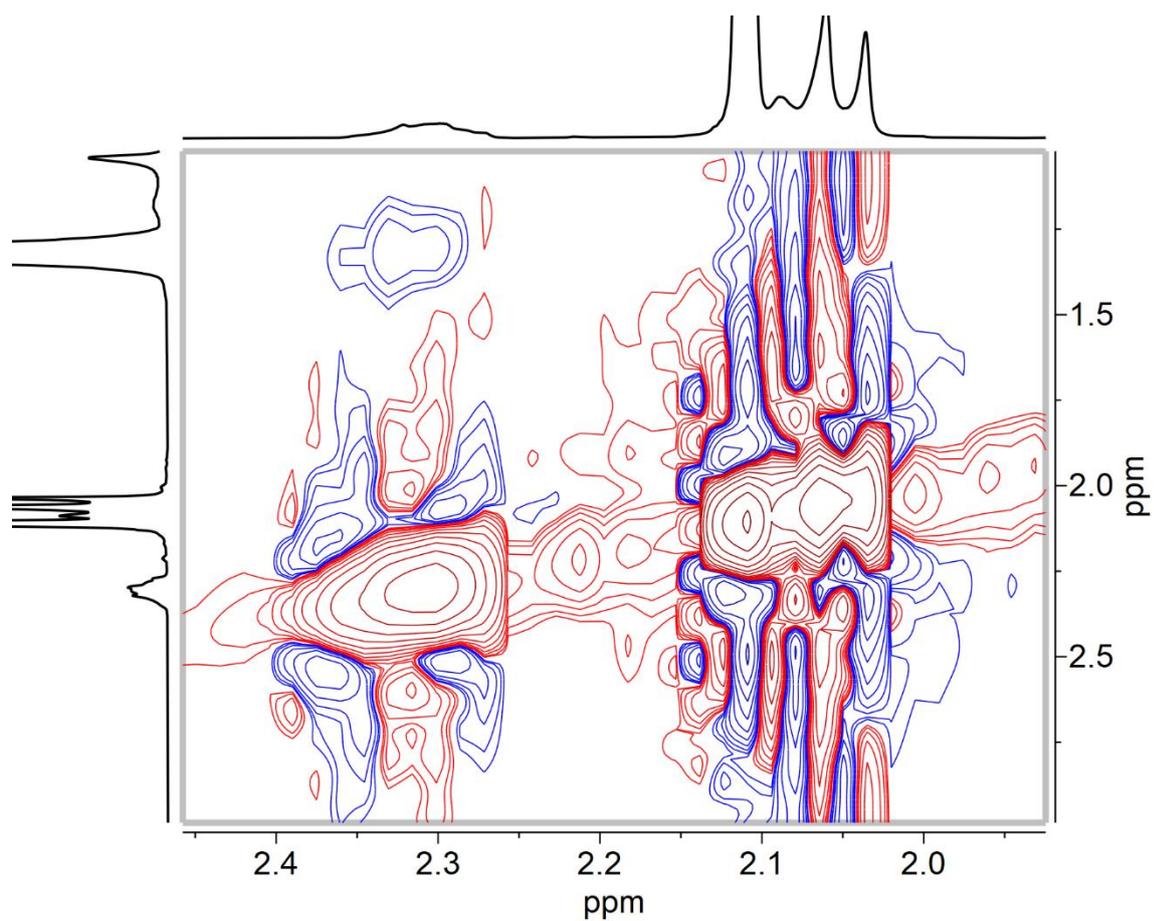

**Figure S15.** $^1$H-$^1$H NOESY NMR spectrum of TTP-Br in toluene-D at 313K, which shows positive (blue) NOE cross peaks, typical of species with the fast-tumbling regime in solution.

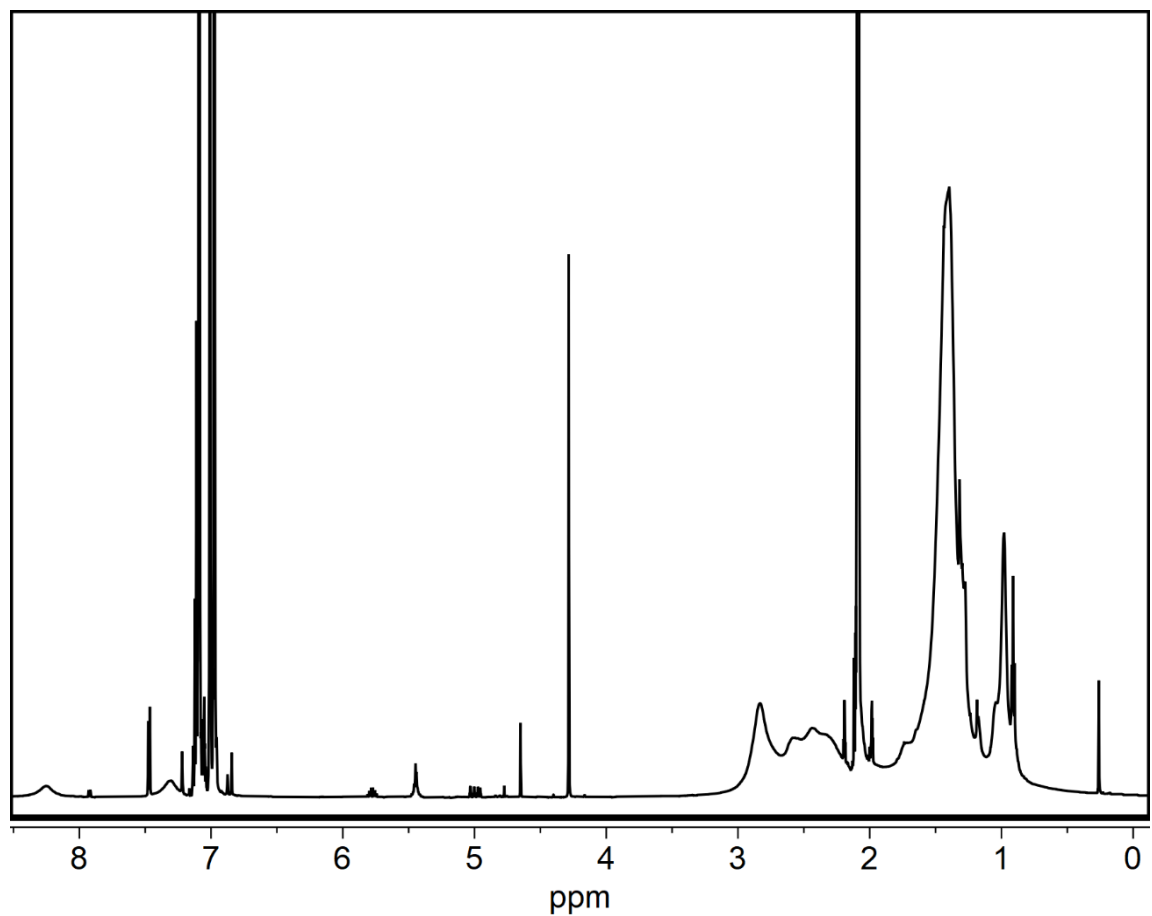

**Figure S16.** $^1$H NMR spectrum of TTP-Br-capped CsPbBr$_3$ NCs in toluene-D at 313K.

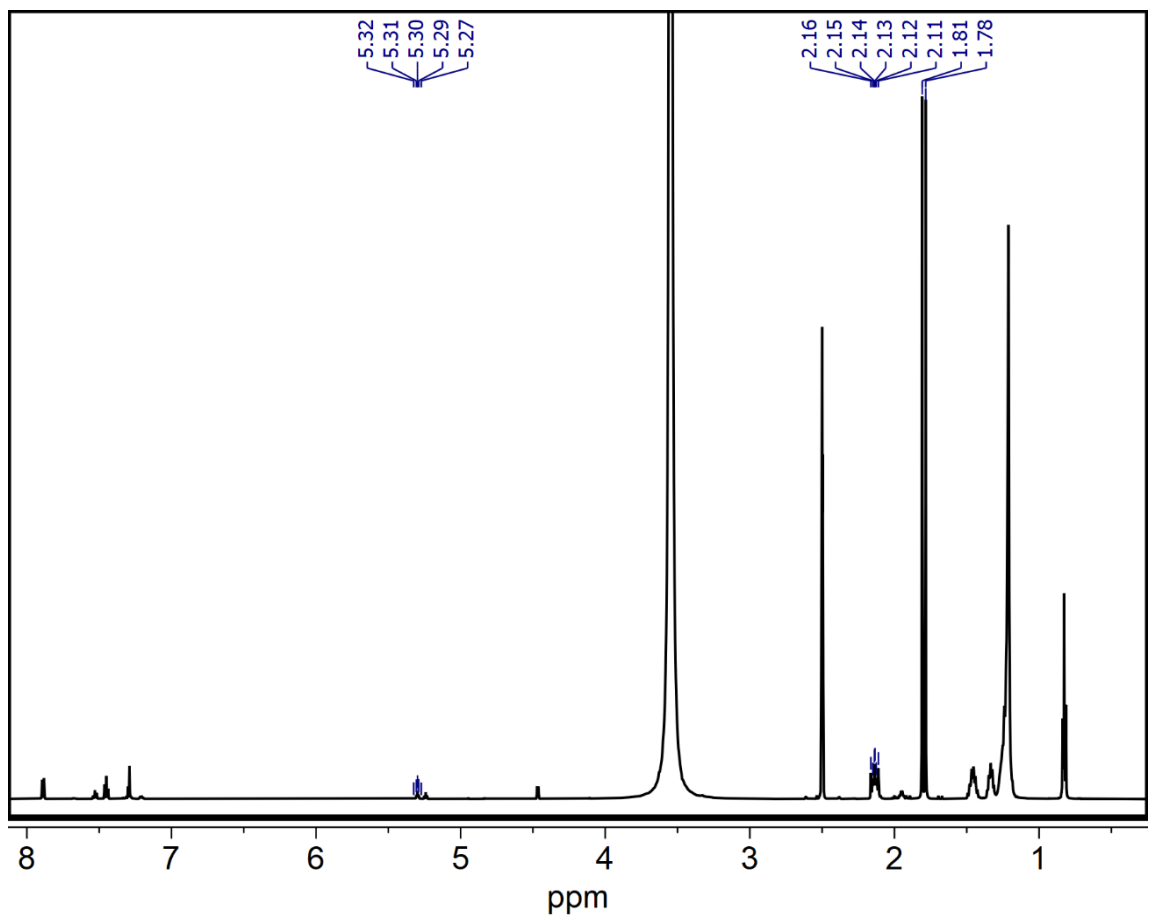

**Figure S17.** *quantitative* ¹H NMR spectrum of TTP-Br ligand in DMSO-*d6*, at 298K, ligand concentration is determined by comparing diagnostic integrated ligand peaks to that of a 10 mM standard dimethyl sulfone solution by using the PULCON method.[7]

**Table S5.** Summary of the parameters employed for the ligand density calculation.

|  | TTP-Br-capped NCs | | DDA-Br-capped NCs | |
|---|---|---|---|---|
|  | **Oleate** | **TTP-Br** | **Oleate** | **DDABr** |
| **[Pb] - ICP-OES (M)** | 1.22E-04 | 1.22E-04 | 3.19E-05 | 3.19E-05 |
| **[Pb]-DMSO (M)** | 2.44E-01 | 2.44E-01 | 6.38E-02 | 6.38E-02 |
| **Nº Pb atoms** | 7.36E+18 | 7.36E+18 | 1.92E+18 | 1.92E+18 |
| **NCs size (nm)** | 8.9 | 8.9 | 8.8 | 8.8 |
| **Unit cell size (nm)** | 5.87E-01 | 5.87E-01 | 5.87E-01 | 5.87E-01 |
| **Unit cells/side** | 1.52E+01 | 1.52E+01 | 1.50E+01 | 1.50E+01 |
| **Nº Pb atoms/ NC** | 3.49E+03 | 3.49E+03 | 3.37E+03 | 3.37E+03 |
| **Nº NCs** | 2.11E+15 | 2.11E+15 | 5.70E+14 | 5.70E+14 |
| **Total NC surface (nm$^2$)** | 1.00E+18 | 1.00E+18 | 2.65E+17 | 2.65E+17 |
| **[Ligands] (M)** | 8.65E-04 | 1.07E-02 | 5.90E-04 | 2.91E-03 |
| **Nº ligand molecules** | 1.04E+17 | 1.29E+18 | 7.11E+16 | 3.50E+17 |
| **Ligands density (Ligands /nm$^2$)** | **0.10** | **1.28** | **0.27** | **1.32** |
| **Surface Ligand %** |  | 92.52 |  | 83.12 |

The Pb and the free ligands concentration were measured by ICP-OES and NMR, respectively, on the samples dissolved in DMSO-d6, whereas the NC size was measured via TEM analysis. The unit cell size was obtained from reference.[13]

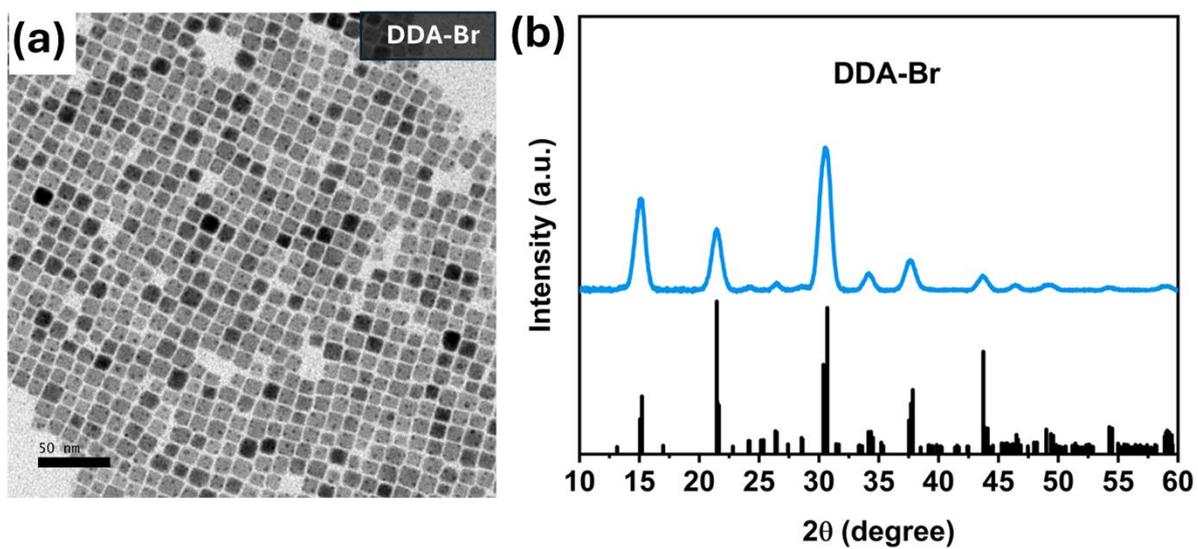

**Figure S18. (a)** TEM micrograph, and (b) XRD spectrum of DDA-Br-capped CsPbBr$_3$ NCs.

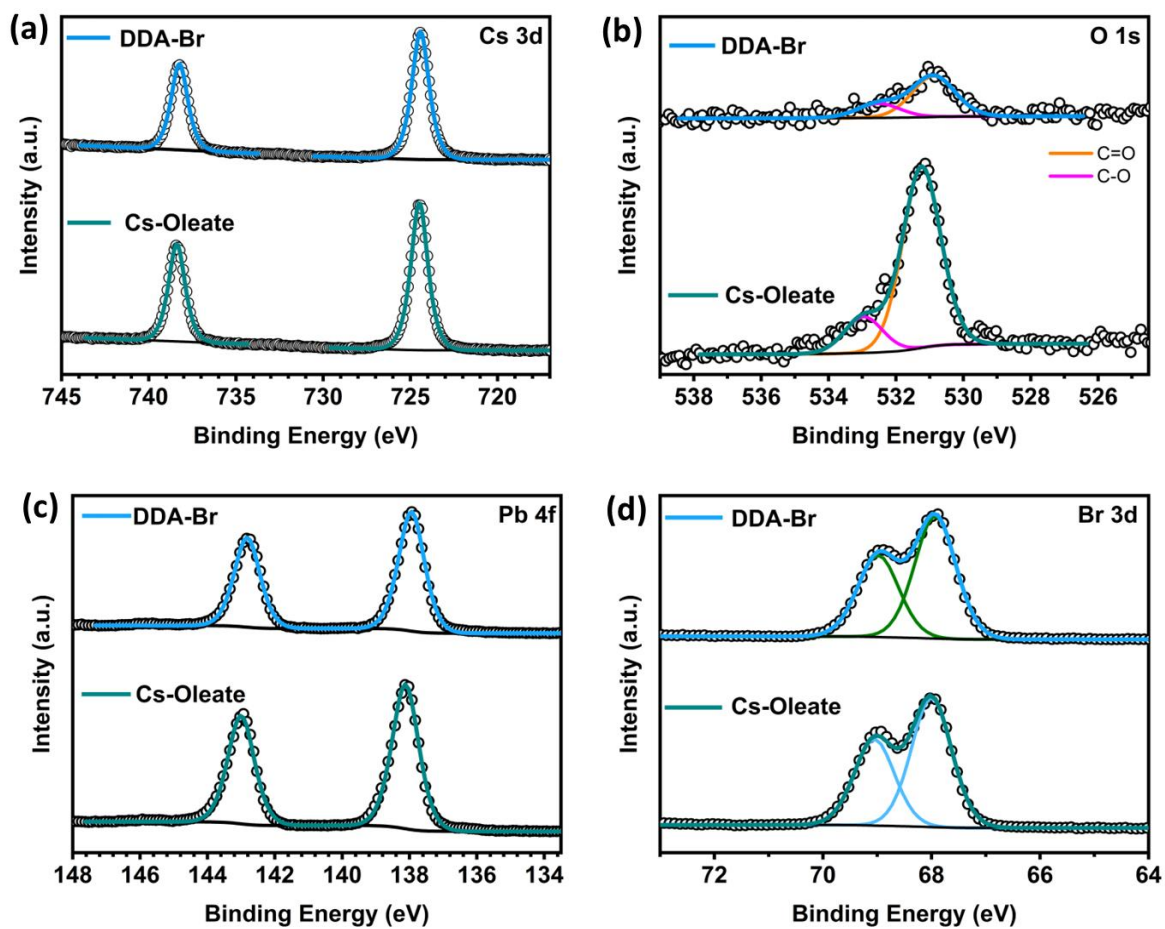

**Figure S19**. Comparison between (a) Cs 3*d*, (b) O 1*s*, (c) Pb 4*f*, (d) Br 3*d*, XPS spectra of DDA-Br-capped, and Cs-Oleate-capped CsPbBr$_3$ NCs.

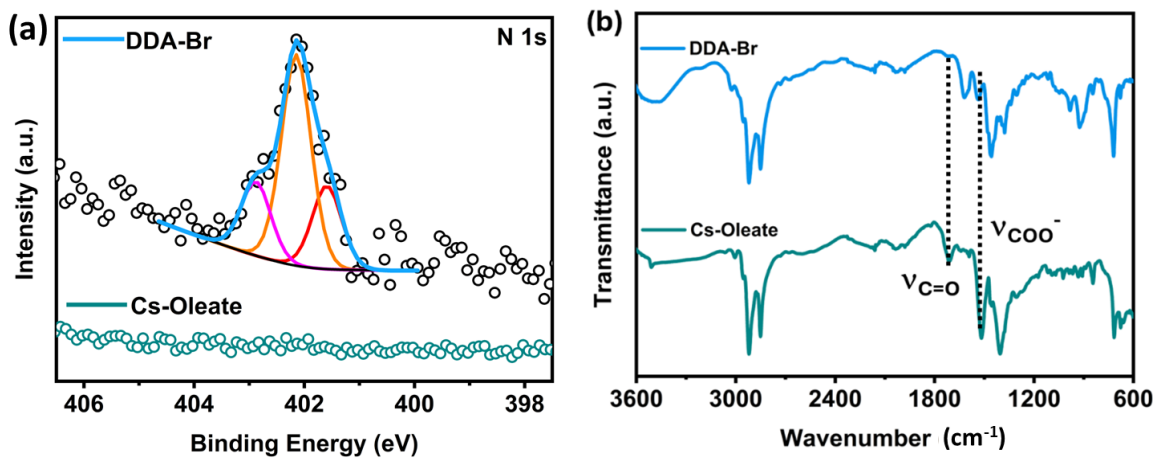

**Figure S20.** (a) Comparison between N 1*s* XPS spectra of DDA-Br- and Cs-Oleate-capped CsPbBr$_3$ NCs, (b) FTIR spectra, of DDA-Br- and Cs-Oleate-capped CsPbBr$_3$ NCs.

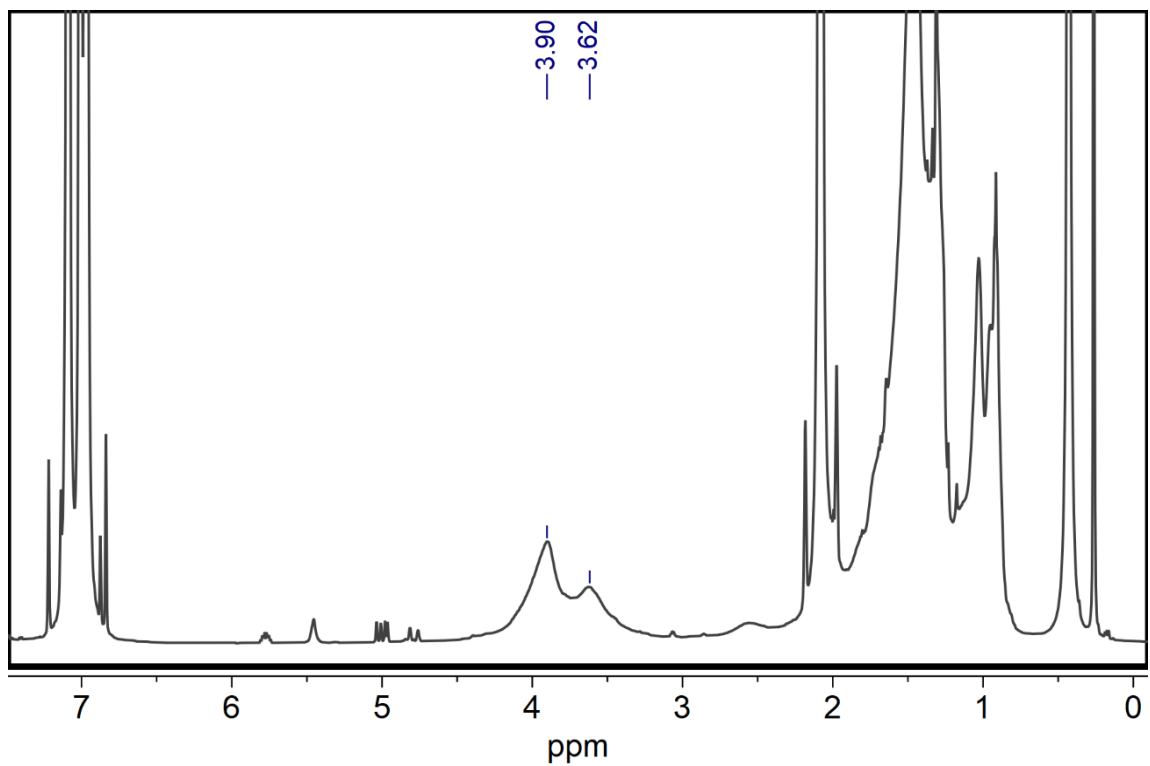

**Figure S21.** $^1$H NMR spectrum of DDA-Br-capped CsPbBr$_3$ NCs in toluene-D at 298K.

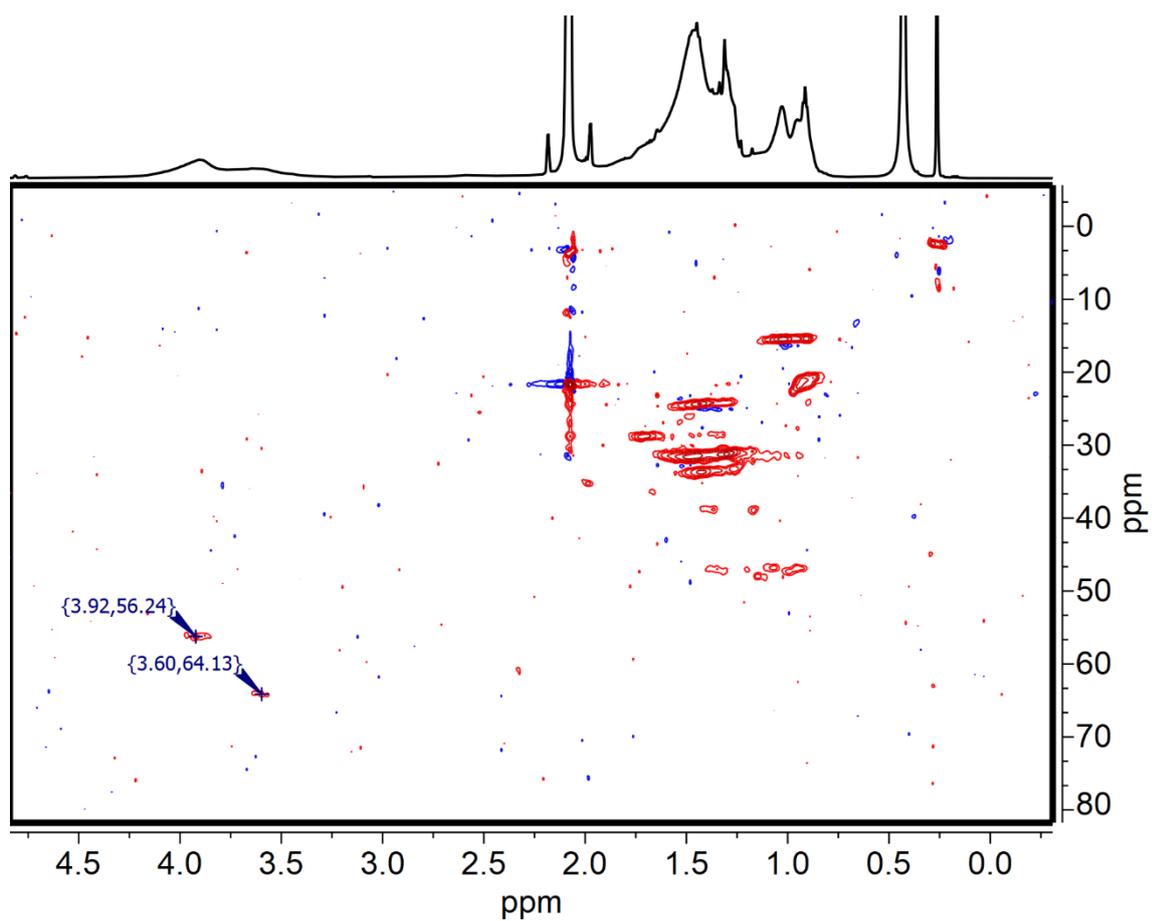

**Figure S22.** $^1$H-$^{13}$C HSQC NMR spectrum of DDA-Br-capped CsPbBr$_3$ NCs in toluene-D at 298K.

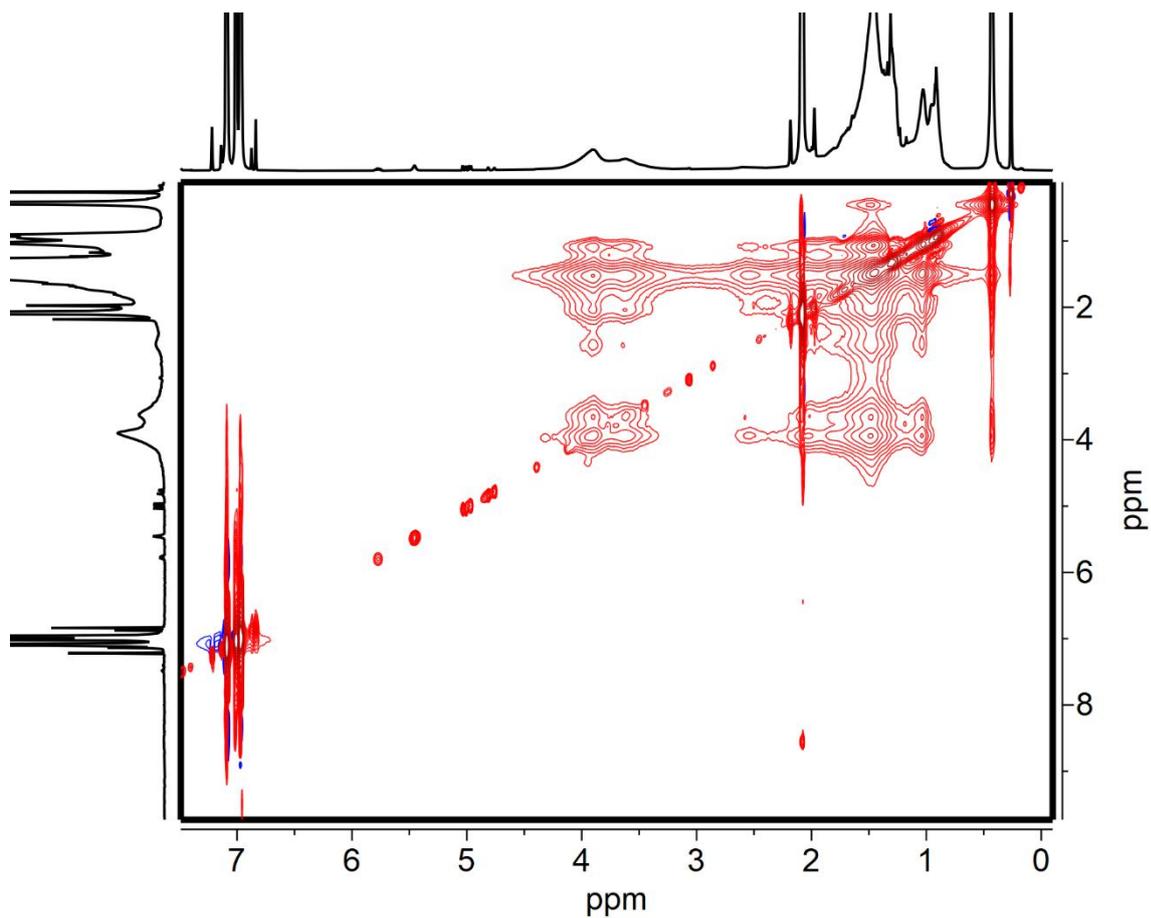

**Figure S23.** $^1$H-$^1$H NOESY NMR spectrum of DDA-Br-capped CsPbBr$_3$ NCs in toluene-D at 298 K, negative (red) NOE cross peaks is typical of species with slow tumbling regime in solution, due to the binding with the NC's surface.

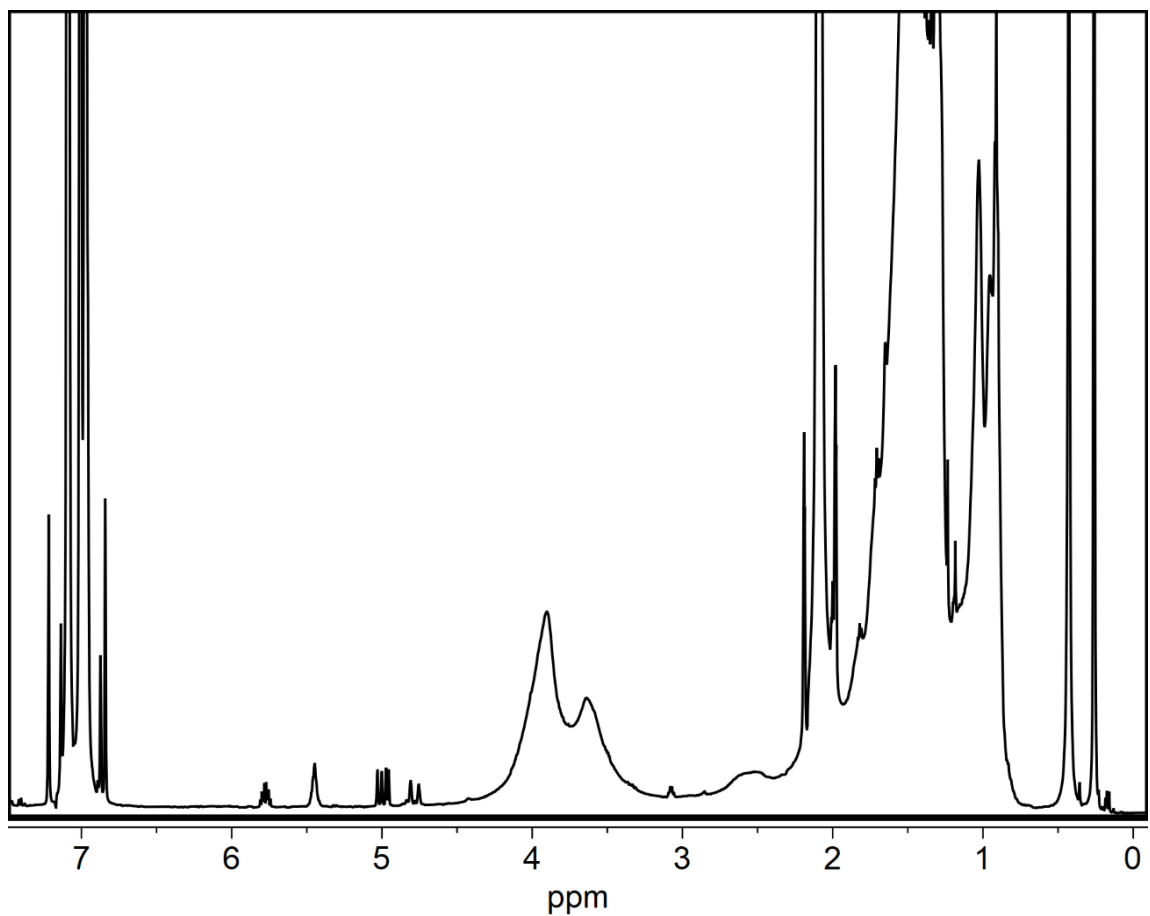

**Figure S24.** $^1$H NMR spectrum of DDA-Br-capped CsPbBr$_3$ NCs in toluene-D at 313K.

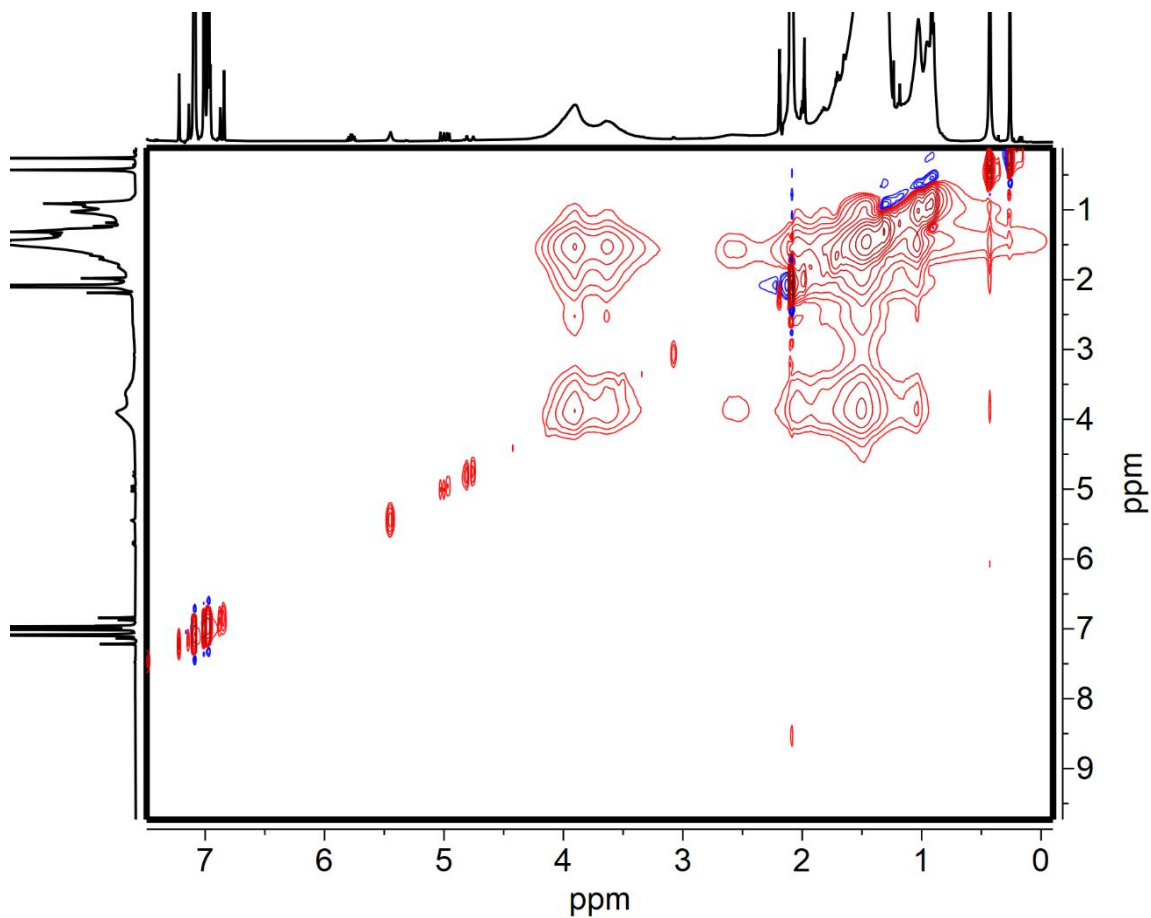

**Figure S25.** $^1$H-$^1$H NOESY NMR spectrum of DDA-Br-capped CsPbBr$_3$ NCs in toluene-D at 313K, negative (red) NOE cross peaks is typical of species with slow tumbling regime in solution, due to the binding with the NC's surface.

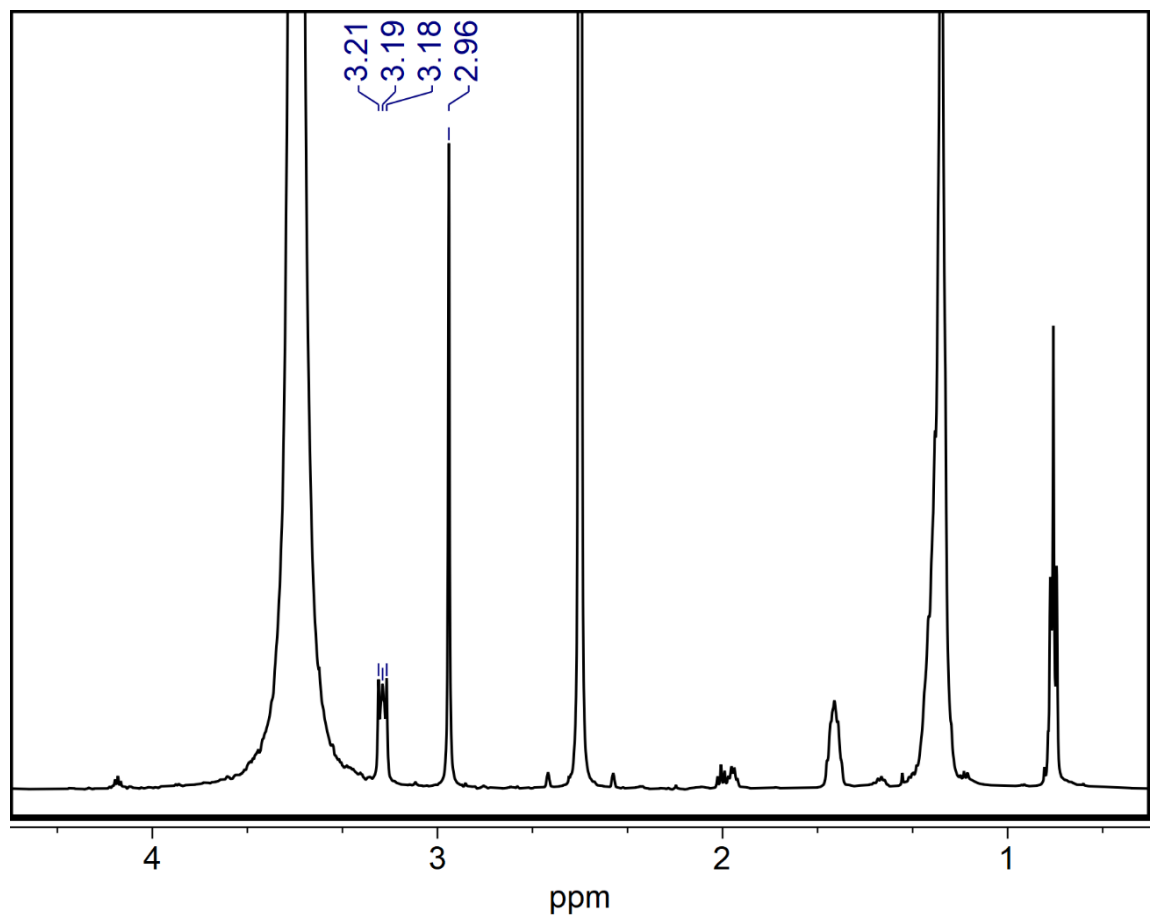

**Figure S26.** *quantitative* $^1$H NMR spectrum of DDA-Br-capped CsPbBr$_3$ NCs in DMSO-*d6*, ligand concentration is determined by comparing diagnostic integrated ligand peaks to that of a 10 mM standard dimethyl sulfone solution, by using PULCON method.[7]

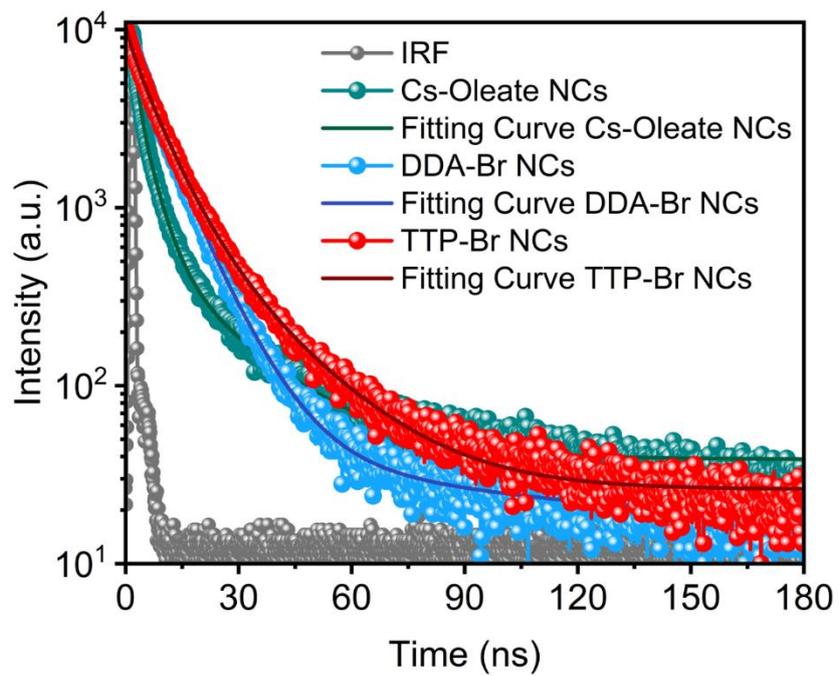

**Figure 27**: PL decay profile with fitting curves of Cs-Oleate-, DDA-Br-, and TTP-Br- capped CsPbBr$_3$ NCs.

**Table: S6.** TRPL fitting parameters and average PL lifetime ($\tau_{ave}$) of the various CsPbBr$_3$ NCs samples.

| NCs Samples | A$_1$ | $\tau_1$ (ns) | A$_2$ | $\tau_2$ (ns) | A$_3$ | $\tau_3$ (ns) | R$^2$ | $\tau_{ave}$ (ns) |
|---|---|---|---|---|---|---|---|---|
| **Cs-Oleate** | 5301 | 1.025 | 4350.8 | 4.13 | 570.01 | 20.05 | 0.99925 | 8.86 |
| **DDA-Br** | 5432.45 | 5.5991 | 4333.18 | 9.9490 | 38.56 | 75.83 | 0.99978 | 10.74 |
| **TTP-Br** | 1268.75 | 1.9233 | 7174.66 | 7.2464 | 1408.88 | 19.68 | 0.99971 | 11.29 |

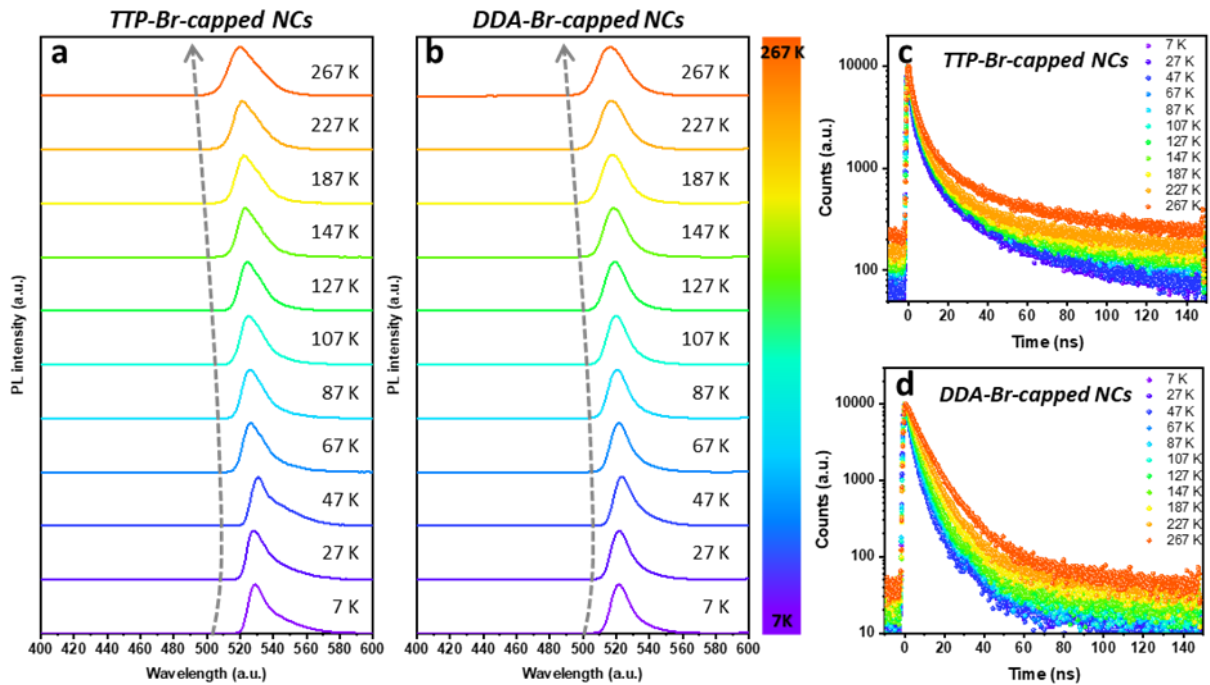

**Figure S28.** PL spectra of (a) TTP-Br-capped and (b) DDA-Br-capped CsPbBr$_3$ NC films recorded from 7 K to 267 K. PL decay traces of (c) TTP-Br-capped and (d) DDA-Br-capped CsPbBr$_3$ NC films recorded from 7 K to 267 K. For low temperature (7 K to 67 K), we are at the time resolution of our instrument.

**Table: S7.** TRPL fitting parameters and average PL lifetime ($\tau_{ave}$ in ns) of TTP-Br-capped CsPbBr$_3$ NCs thin film recorded at different temperatures.

| TTP-Br-capped CsPbBr$_3$ NCs Biexponential decay curve fitting | | | | | | |
|---|---|---|---|---|---|---|
| Temperature (K) | A$_1$ | $\tau_1$ (ns) | A$_2$ | $\tau_2$ (ns) | R$^2$ | $\tau_{Average}$ (ns) |
| 7 | 8095.12 | 1.31 | 2470.02 | 11.76 | 0.99562 | 8.97 |
| 27 | 7930.40 | 1.34 | 2354.04 | 11.67 | 0.99592 | 8.79 |
| 47 | 7868.33 | 1.40 | 2412.78 | 12.34 | 0.99562 | 9.39 |
| 67 | 7555.12 | 1.61 | 2624.13 | 12.54 | 0.99638 | 9.59 |
| 87 | 7736.04 | 1.64 | 2725.74 | 12.16 | 0.99658 | 9.25 |
| 107 | 7545.79 | 1.77 | 2751.27 | 12.33 | 0.99696 | 9.34 |
| 127 | 7697.44 | 1.73 | 2858.17 | 11.33 | 0.99685 | 8.53 |
| 147 | 7618.06 | 1.91 | 2798.72 | 11.86 | 0.99753 | 8.83 |
| 187 | 7620.62 | 2.09 | 2716.30 | 12.30 | 0.99757 | 9.01 |
| 227 | 7706.80 | 2.17 | 2656.23 | 12.94 | 0.99758 | 9.41 |
| 267 | 7682.56 | 2.56 | 2470.87 | 17.22 | 0.99736 | 12.58 |

**Table: S8.** TRPL fitting parameters and average PL lifetime ($\tau_{ave}$ in ns) of DDA-Br-capped CsPbBr$_3$ NCs thin film recorded at different temperatures.

| DDA-Br-capped CsPbBr$_3$ NCs Biexponential decay curve fitting | | | | | | |
|---|---|---|---|---|---|---|
| Temperature (K) | A$_1$ | $\tau_1$ (ns) | A$_2$ | $\tau_2$ (ns) | R$^2$ | $\tau_{Average}$ (ns) |
| 7 | 9877.25 | 2.91 | 1165.97 | 11.05 | 0.99969 | 5.43 |
| 27 | 9793.64 | 2.80 | 1286.31 | 10.37 | 0.99965 | 5.28 |
| 47 | 9887.97 | 2.71 | 1161.89 | 10.01 | 0.99975 | 4.92 |
| 67 | 9672.57 | 2.92 | 1455.09 | 10.34 | 0.9996 | 5.50 |
| 87 | 9679.35 | 3.25 | 1318.01 | 11.31 | 0.99942 | 5.84 |
| 107 | 9554.99 | 3.55 | 1335.15 | 11.44 | 0.99963 | 6.00 |
| 127 | 9832.43 | 3.88 | 1200.08 | 12.13 | 0.99942 | 6.16 |
| 147 | 9952.69 | 4.28 | 1007.83 | 13.54 | 0.99952 | 6.53 |
| 187 | 10368.92 | 5.42 | 546.58 | 19.79 | 0.99934 | 7.74 |
| 227 | 10302.02 | 6.09 | 617.28 | 20.99 | 0.99938 | 8.64 |
| 267 | 9947.19 | 7.30 | 878.38 | 22.30 | 0.99939 | 10.49 |

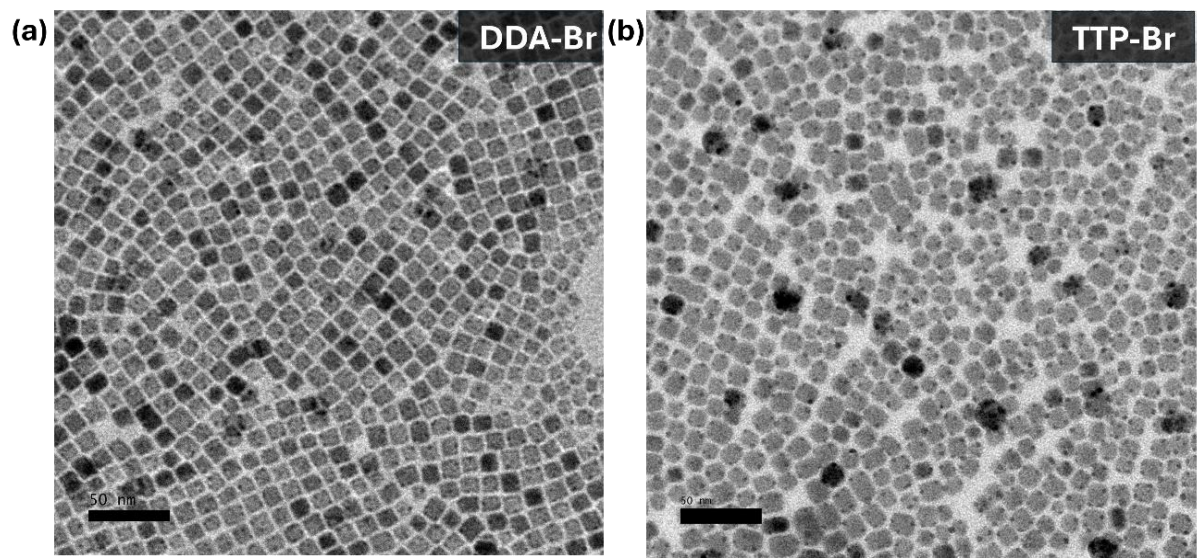

**Figure S29.** TEM micrographs of (a) DDA-Br-capped CsPbBr$_3$ NCs (average size of 10.3 ± 1.6 nm) and (b) TTP-Br-capped CsPbBr$_3$ NCs (average size of 9.8 ± 1.8 nm) after six weeks of storage under air.

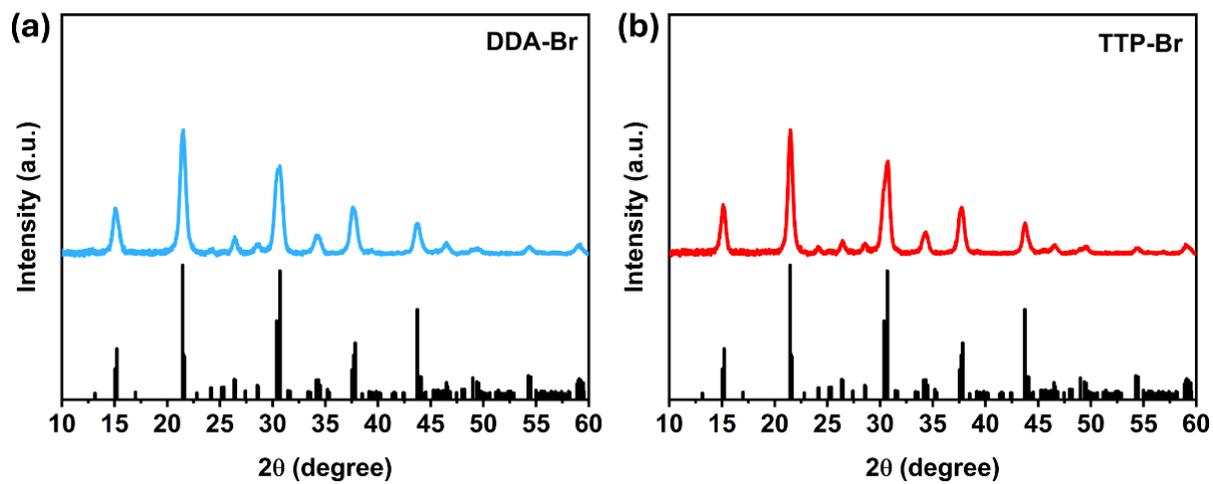

**Figure S30.** XRD spectra of (a) DDA-Br-capped, (b) TTP-Br-capped $CsPbBr_3$ NCs after six weeks of storage under air.

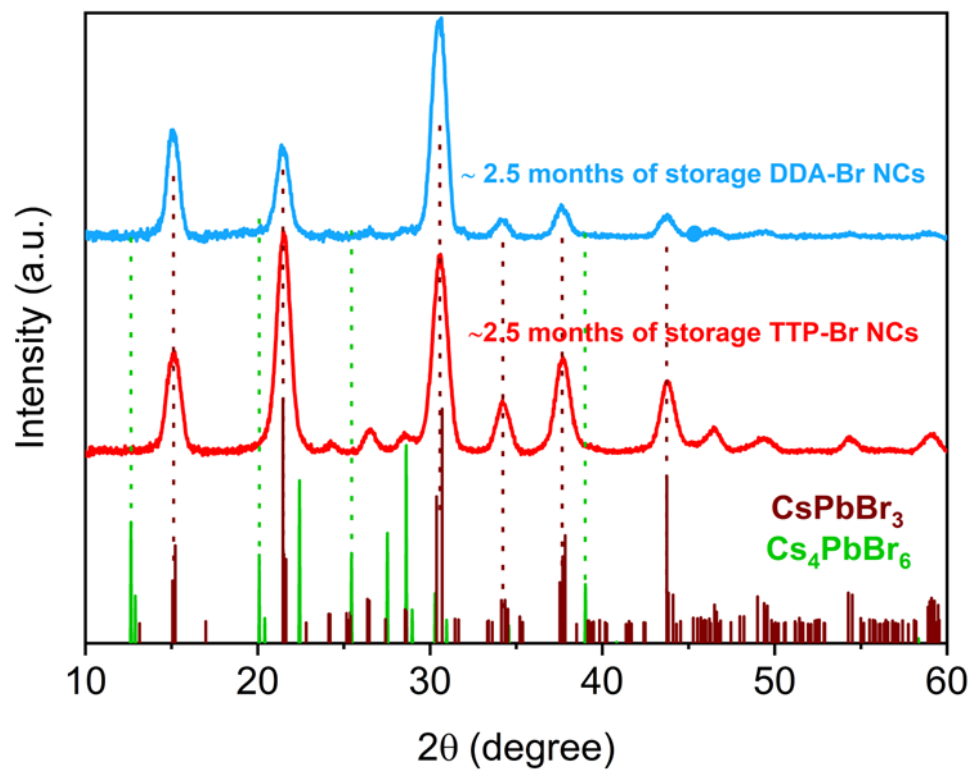

**Figure S31**: XRD spectra of TTP-Br-capped and DDA-Br-capped CsPbBr$_3$ NCs after storage of ~2.5 months at ambient air conditions.

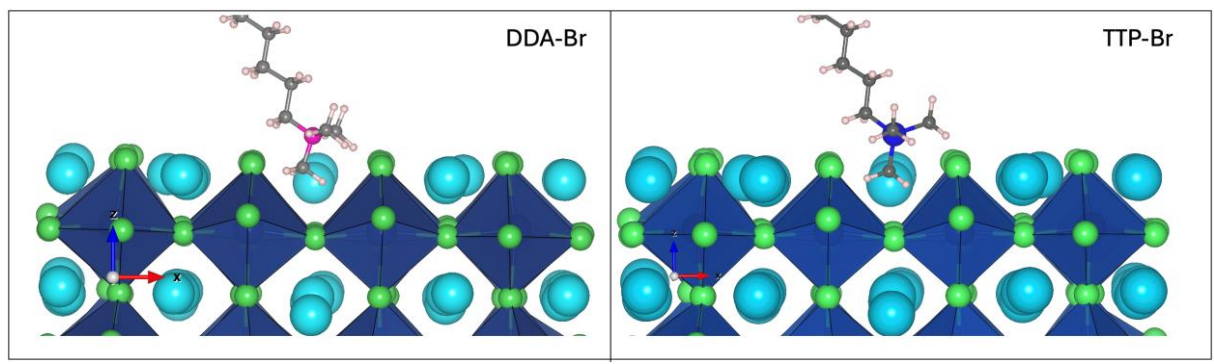

**Figure S32.** Binding configuration of (left) a DDA$^+$ and (right) a TTP$^+$ ligands sitting in the A-site of the CsPbBr$_3$ NCs' surface after structural relaxation at the DFT/PBE level. The quaternary ammonium and phosphonium ligands assume similar orientations, with one of the N-CH$_3$ or P-CH$_3$ bonds almost perpendicular to the NC surface.

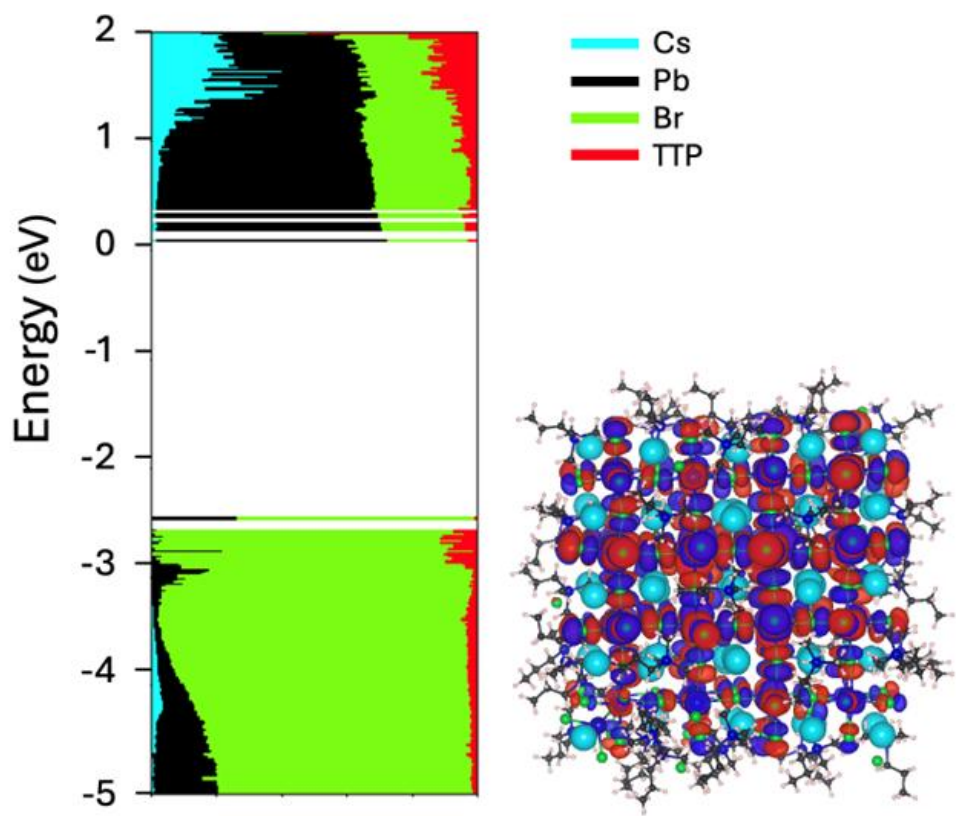

**Figure S33.** (left) Electronic structure of the ~2.4 nm-sided CsPbBr$_3$ NC model passivated with TTP-Br ligand with a surface concentration of 1.27 ligands/nm$^2$ computed at the DFT/PBE level of theory. The color code indicates the contribution of each atom type to each molecular orbital and the C, H and P contributions from TTP-Br ligands are grouped for clarity. (right) Isosurface of the valence band edge orbital with a counter value of 0.02 e/Bohr$^3$.

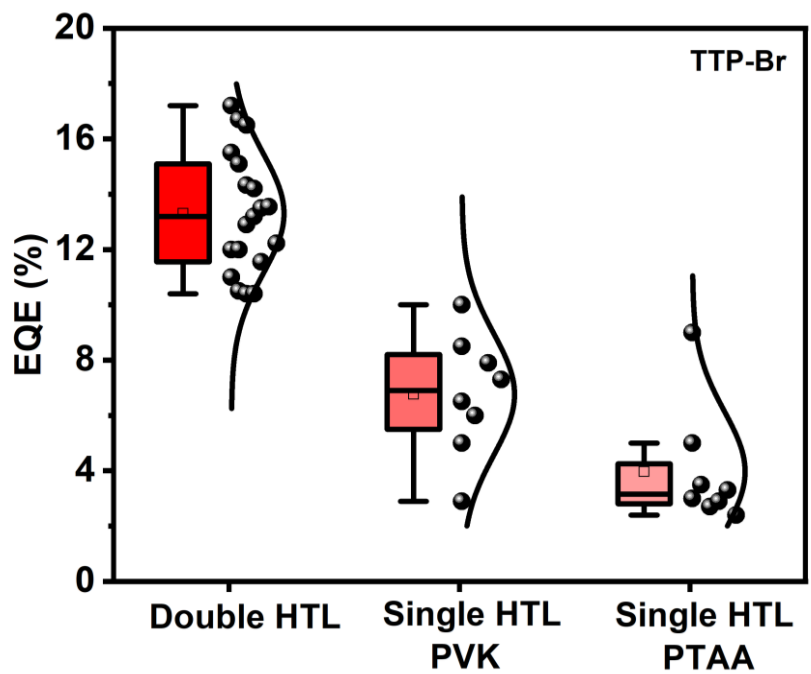

**Figure S34**. Statistics of EQE values of TTP-Br-capped $CsPbBr_3$ NCs-based LEDs having either a double or single HTL configuration.

**Table S9.** UPS analysis of CsPbBr$_3$ NCs.

| CsPbBr$_3$ NCs | WF | E$_{ion}$ | E$_1$ = E$_{ion}$ - WF |
|---|---|---|---|
| **TTP-Br** | 4.52 | 5.31 | 0.79 |
| **DDA-Br** | 4.57 | 5.30 | 0.73 |

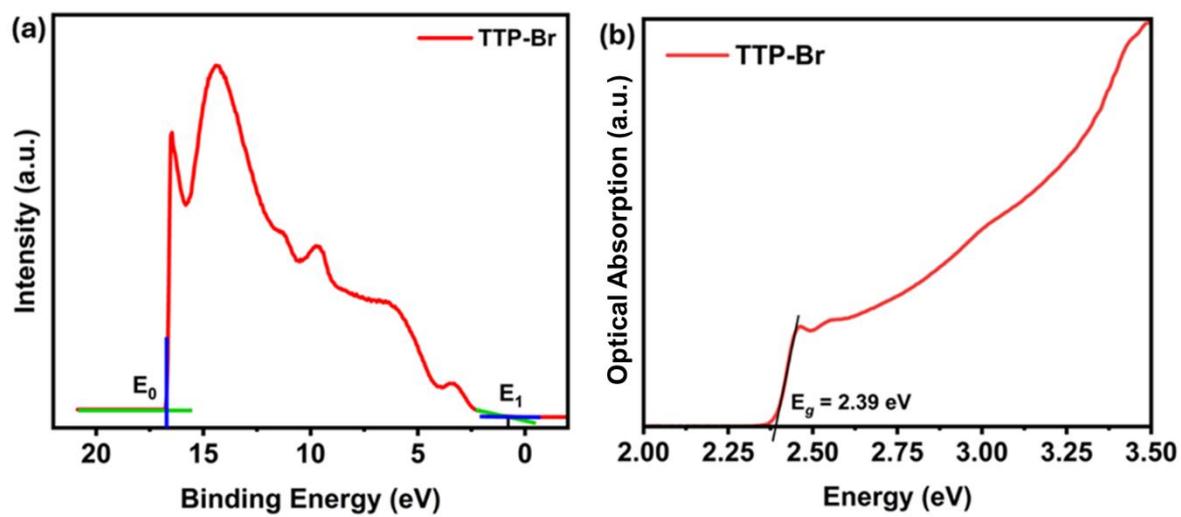

**Figure S35.** (a) UPS spectrum, (b) absorption spectrum of TTP-Br-capped $CsPbBr_3$ NCs.

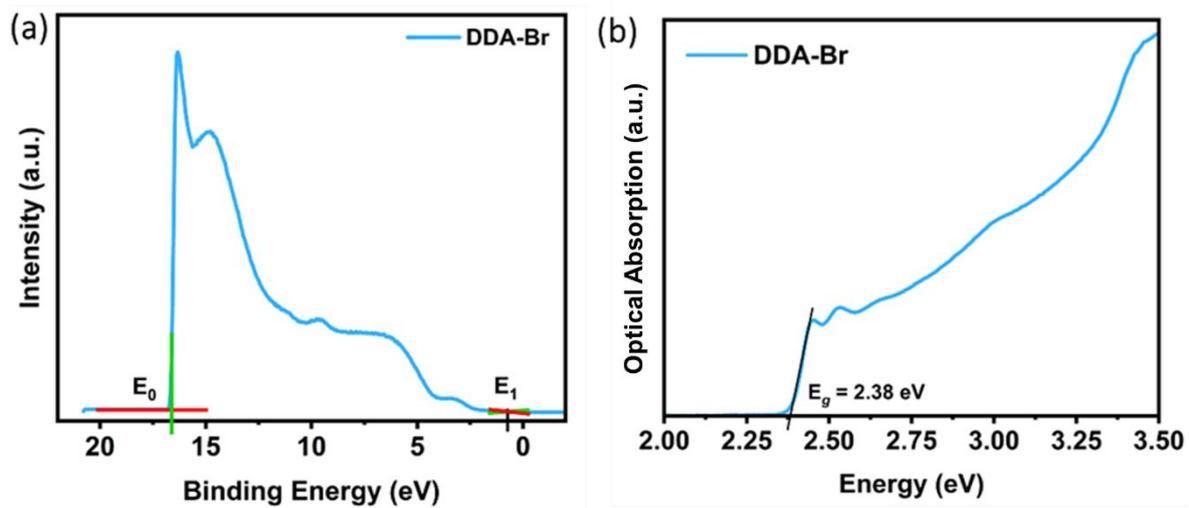

**Figure S36.** (a) UPS spectrum, (b) absorption spectrum of DDA-Br-capped CsPbBr$_3$ NCs.

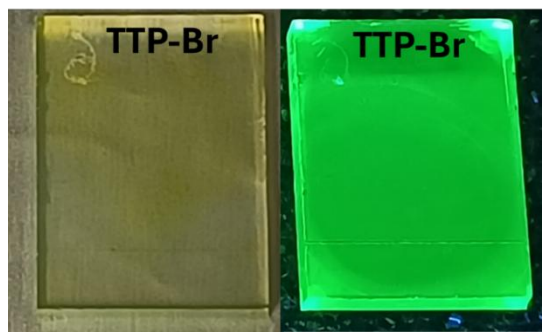

**Figure S37.** TTP-Br-capped CsPbBr$_3$ NC films under normal light and under UV light.

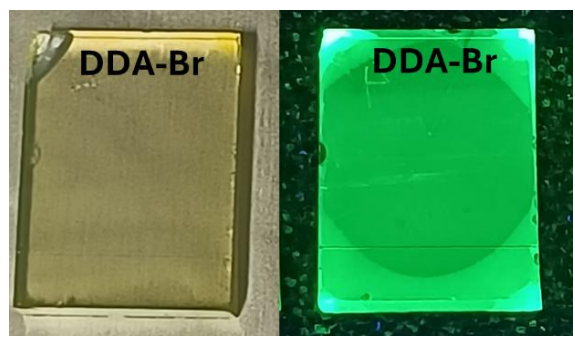

**Figure S38.** DDA-Br-capped CsPbBr$_3$ NC films under normal light and under UV light.

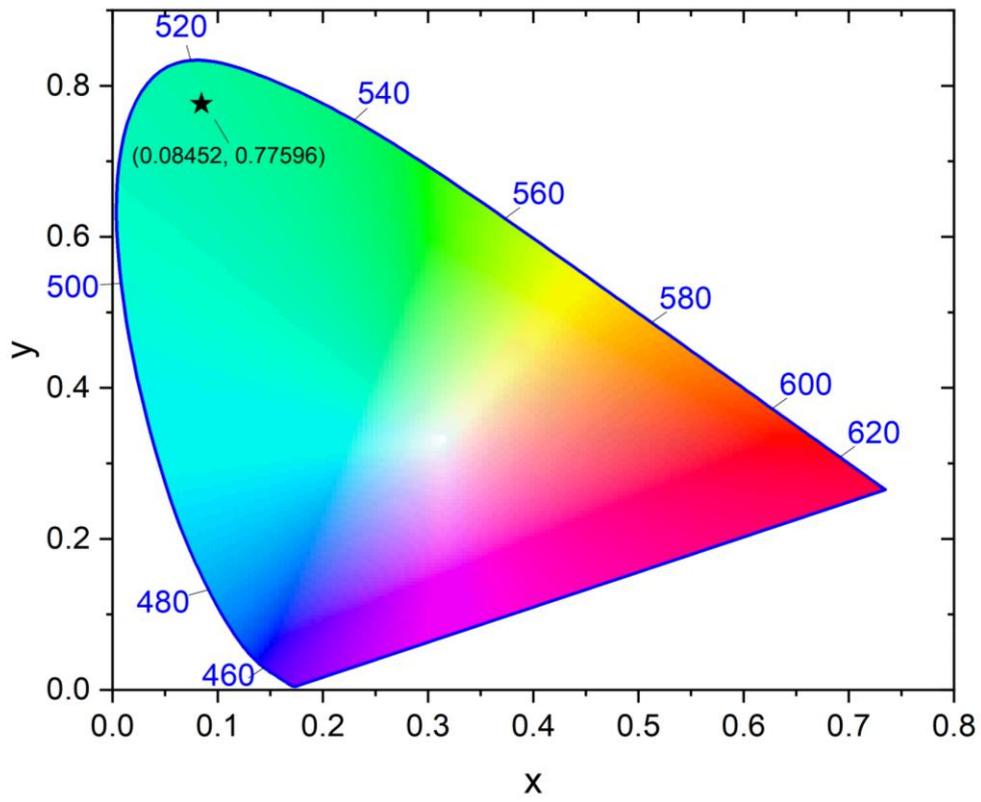

**Figure S39**. Commission Internationale de l'Eclairage (CIE) 1931 colour coordinates.

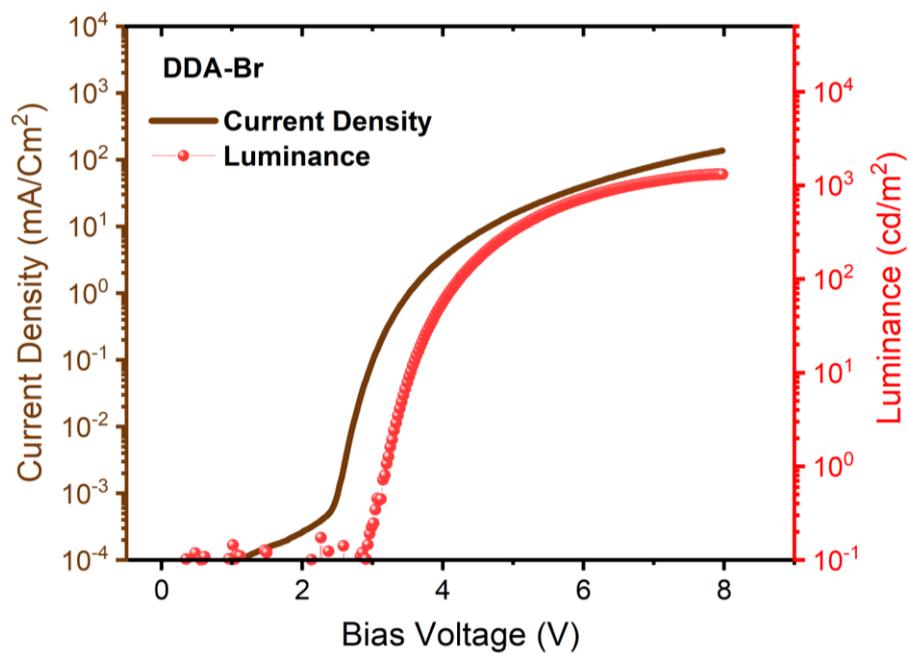

**Figure S40**. Current density and luminescence versus driving voltage curves of the DDA-Br-capped CsPbBr$_3$ NCs LED with double HTL.

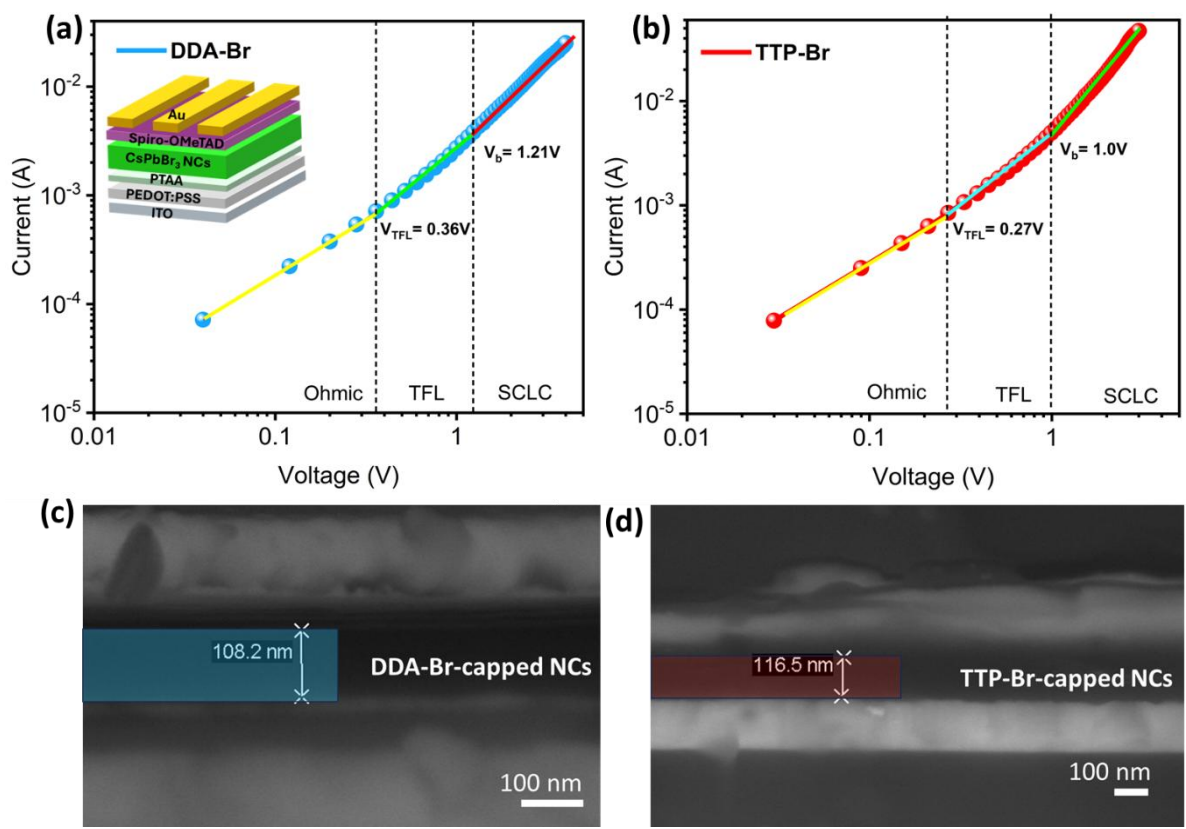

**Figure S41.** Space charge limit current (SCLC) measurements of the hole-only devices, with the structure (a) DDA-Br-capped, (b) TTP-Br-capped CsPbBr$_3$ NCs. Cross-sectional SEM images of the thin films with (c) DDA-Br-capped, and (d) TTP-Br-capped CsPbBr$_3$ NCs.

The hole mobility was determined by the Mott-Gurney equation[14]:

$$J_D = \frac{9\varepsilon\varepsilon_0\mu V_b^2}{8L^3}$$

where ε is the relative permittivity of the CsPbBr$_3$ NCs (4.8),[15] $\varepsilon_0$ is the vacuum permittivity (8.85 x 10$^{-12}$ F/m), $V_{TFL}$ is the trap-filled limit voltage, $V_b$ is the voltage at which the charge transport transitions from a trap-filling regime to an SCLC regime, $J_D$ is the current density at $V_b$ point, $\mu$ is the charge carrier mobility, and L is the thickness of the CsPbBr$_3$ NC film (L = 108.2 nm for DDA-Br-capped CsPbBr$_3$ NCs, and L = 116.5 nm for TTP-Br capped CsPbBr$_3$ NCs respectively).

**Table S10.** Hole carrier mobility of the DDA-Br-capped, and TTP-Br-capped CsPbBr$_3$ NC films from the fits of the SCLC curves.

| CsPbBr$_3$ NCs | $V_{TFL}$ (V) | $V_b$ (V) | L (nm) | $\mu$ (cm$^2$V$^{-1}$s$^{-1}$) |
|---|---|---|---|---|
| **DDA-Br** | 0.36 | 1.21 | 108.2 | 6.87 x 10$^{-7}$ |
| **TTP-Br** | 0.27 | 1.0 | 116.5 | 1.43 x 10$^{-6}$ |

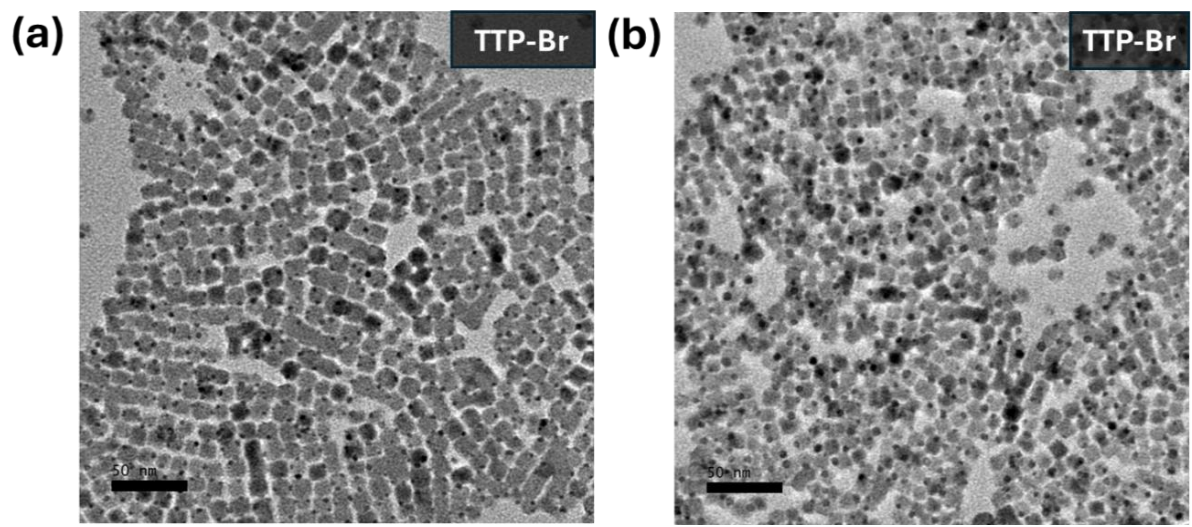

**Figure S42.** TEM micrographs of TTP-Br-capped CsPbBr$_3$ NCs, (a) after the second treatment (2.5 mM) followed by washing, and (b) after the third treatment (2.5 mM) followed by washing.